%% file: main.tex
\pdfoutput=1
\documentclass[journal,10pt,twocolumn,oneside]{IEEEtran}

\usepackage{booktabs}
\usepackage{wrapfig}
\usepackage{verbatim}
\usepackage{algorithm}
\usepackage{algorithmicx}
\usepackage{algpseudocode}
\usepackage{graphicx}
\usepackage{cite}
\usepackage{hyperref}
\usepackage[cmex10]{amsmath}
\usepackage{amsthm}
\usepackage{epstopdf}
\usepackage{caption}
\usepackage{subcaption}

\usepackage{makecell}
\usepackage{arydshln}
\usepackage{multirow}
\usepackage{multicol}
\usepackage{nicefrac}
\usepackage{cuted}
\usepackage{flushend}
\let\txtb\b \let\txtc\c \let\txtd\d \let\txtH\H \let\txti\i \let\txtj\j
\let\txtk\k \let\txtl\l \let\txtL\L \let\txto\o \let\txtO\O \let\txtP\P
\let\txtr\r \let\txtS\S \let\txtt\t \let\txtu\u \let\txtv\v
\newcommand{\restoretextaccents}{%
  \let\b\txtb \let\c\txtc \let\d\txtd \let\H\txtH \let\i\txti \let\j\txtj
  \let\k\txtk \let\l\txtl \let\L\txtL \let\o\txto \let\O\txtO \let\P\txtP
  \let\r\txtr \let\S\txtS \let\t\txtt \let\u\txtu \let\v\txtv}

\input{defs}

\usepackage{color}

\usepackage{mdwmath}
\usepackage{url}
\usepackage{amssymb}
\usepackage{pgfplots}
\pgfplotsset{compat=newest}

\def\ie{{\textit{i.e.}}}

\newtheorem{theorem}{\textbf{Theorem}}

\newtheorem{corollary}{\textbf{Corollary}}

\begin{document}

\title{Joint Image Denoising and Interpolation via \\
Mixed Graph Filtering}
\author{%
\IEEEauthorblockN{Gene Cheung, \emph{Fellow, IEEE}, Niruhan Viswarupan, \emph{Student Member, IEEE}, Romisaa Elsaidy}
\renewcommand{\baselinestretch}{1.0}
\thanks{The work of G. Cheung was supported in part by the Natural Sciences and Engineering Research
Council of Canada (NSERC) RGPIN-2025-06252. \emph{(Corresponding author: Gene Cheung.)}}
\thanks{{G. Cheung, N. Viswarupan and Romisaa Elsaidy are with the department of EECS, York University, 4700 Keele Street, Toronto, M3J 1P3, Canada (e-mail:genec@yorku.ca, niruhanv@gmail.com, romaisaa@yorku.ca).}}
}

\markboth{Submitted to IEEE Transactions on Signal Processing}%
{Shell \MakeLowercase{\textit{et al.}}: A Sample Article Using IEEEtran.cls for IEEE Journals}

\maketitle
\begin{abstract}
\input{abstract}
\end{abstract}

\maketitle

\section{Introduction}
\input{intro}

\graphicspath{{fig/}}

\section{Related works}
\label{sec:related}

\input{related}

\section{Preliminaries}
\label{sec:prelim}
\input{prelim.tex}

\section{Graph Interpretation of Interpolator}
\label{sec:graphInterpolate}
\input{graphInterpolate}

\section{Joint Denoising / Interpolation Optimization}
\label{sec:JDI}
\input{JDI}

\section{Generalization to Multiple Interpolators}
\label{sec:generalInterpolate}
\input{interpolate}

\section{Experimentation}
\label{sec:results}
\input{results}

\section{Conclusion}
\label{sec:conclude}
\input{conclude}

\bibliographystyle{IEEEtran}
{\restoretextaccents
\bibliography{ref2}}

\end{document}

%% file: defs.tex
\def\0{{\mathbf 0}}
\def\1{{\mathbf 1}}

\def\b{{\mathbf b}}
\def\c{{\mathbf c}}
\def\d{{\mathbf d}}
\def\e{{\mathbf e}}

\def\i{{\mathbf i}}
\def\j{{\mathbf j}}
\def\k{{\mathbf k}}
\def\l{{\mathbf l}}

\def\o{{\mathbf o}}
\def\p{{\mathbf p}}

\def\r{{\mathbf r}}

\def\t{{\mathbf t}}
\def\u{{\mathbf u}}
\def\v{{\mathbf v}}

\def\x{{\mathbf x}}
\def\y{{\mathbf y}}

\def\A{{\mathbf A}}
\def\B{{\mathbf B}}
\def\C{{\mathbf C}}
\def\D{{\mathbf D}}

\def\G{{\mathbf G}}

\def\H{{\mathbf H}}
\def\I{{\mathbf I}}

\def\L{{\mathbf L}}

\def\O{{\mathbf O}}
\def\P{{\mathbf P}}
\def\Q{{\mathbf Q}}

\def\S{{\mathbf S}}

\def\V{{\mathbf V}}

\def\ie{{\textit{i.e.}}}

\def\cE{{\mathcal E}}

\def\cG{{\mathcal G}}

\def\cL{{\mathcal L}}

\def\cV{{\mathcal V}}

\def\cO{{\mathcal O}}

\def\bPsi{{\boldsymbol \Psi}}

\def\bTheta{{\boldsymbol \Theta}}

%% file: abstract.tex
Images often require both denoising and interpolation, yet treating the two tasks separately is suboptimal: 
interpolation introduces correlated noise that complicates subsequent denoising, while denoising may remove image details useful for interpolation. 
In this paper, we study joint image denoising and interpolation from a mixed graph filtering perspective, where denoising is modeled using undirected graph filters and interpolation using directed graph filters. 
We first establish that, under mild conditions, any (pseudo-)linear denoiser and interpolator can be interpreted as solution filters of maximum a posteriori (MAP) estimation problems regularized by undirected and directed graph smoothness priors, respectively. 
Guided by these results, we formulate graph-based joint denoising and interpolation problems using both directed and undirected graph structures. 
We characterize when joint denoising/interpolation reduces to a separable pipeline and when a genuinely coupled optimization is required.
For the non-separable case, we recover sparse graph operators from given denoisers and interpolators, yielding an efficient linear-system formulation solvable via conjugate gradient. 
Experimental results demonstrate that our proposed joint framework consistently outperforms separate denoising and interpolation pipelines across multiple denoiser-interpolator combinations.

%% file: intro.tex
2D images acquired from physical sensors are typically noise-corrupted, and often a subsequent interpolation task is required for processing and/or display purposes. 
For example, an image acquired on a Bayer-patterned grid, where the sensor obtains only one color of three (Red, Green, and Blue) for each pixel, requires \textit{demosaicking} \cite{Hirakawa2007}: color interpolation so that each pixel possesses all three colors. 

Given the need for both image denoising and interpolation, one naive approach is to first interpolate and then denoise in separate steps; since each additive pixel noise is spread to neighboring pixels during interpolation, the latter denoising task is made more difficult when faced with correlated noise.
On the other hand, first denoising an image---before interpolation like demosaicking that smear noise in a local area---can improve performance \cite{Park2009}. 
However, doing so also means that image details unintentionally lost during denoising can no longer be utilized by the interpolator.

In light of the advent of deep neural networks (DNN) such as convolutional neural nets (CNN) \cite{lecun1998gradient} and transformers \cite{Vaswani2017}, there exist joint image denoising / interpolation methods that learn mappings directly from noisy pixels to interpolated pixels in a data-driven manner \cite{Jin2020,Liu2020,Xing2021}.  
However, such methods learn task-specific mappings from training data and often require retraining or adaptation when sensing conditions or interpolation objectives change.
Given the maturity of existing image denoisers and interpolators, rather than developing yet another denoiser or interpolator, we ask a practical question: \textit{given an existing image denoiser $\bPsi$ and an image interpolator $\bTheta$, how to combine them in a mathematically optimal way into joint image denoising / interpolation operations?}

\begin{figure}[t]
    \centering
    \includegraphics[width=\linewidth]{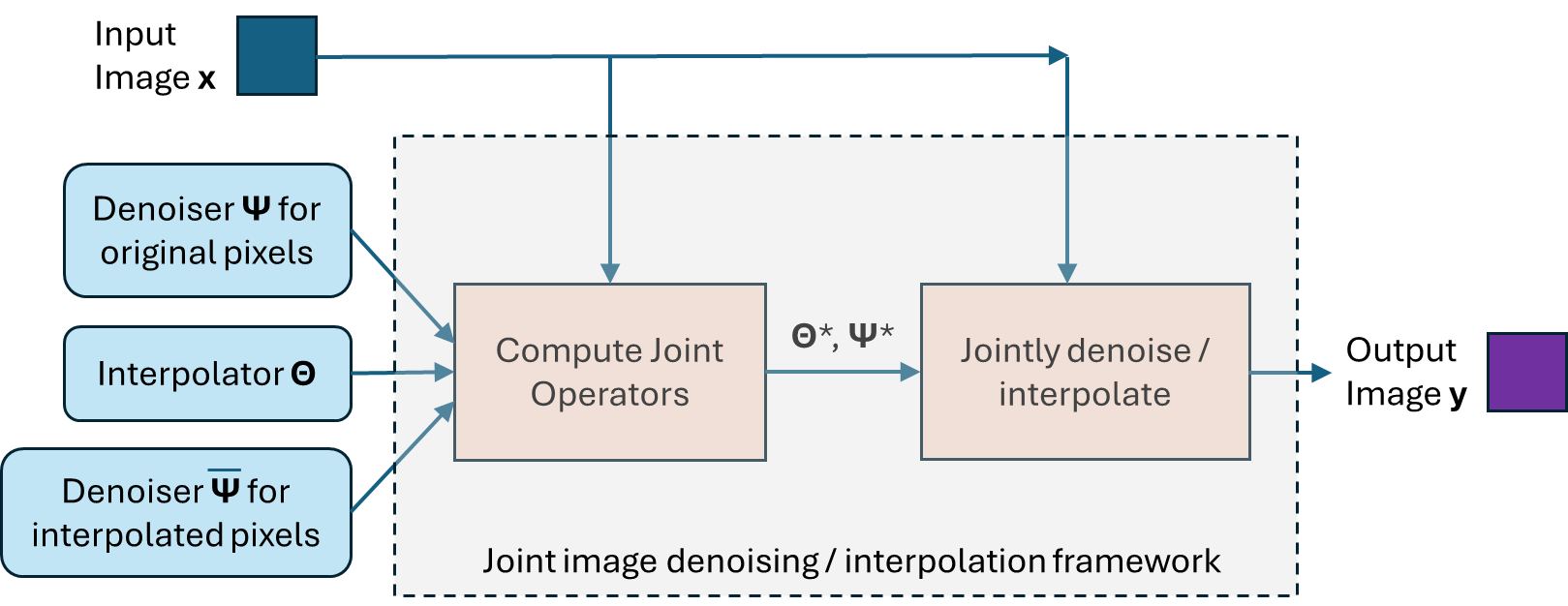}
    \caption{Overview of proposed joint image denoising / interpolation framework. Given (pseudo-)linear denoisers $\bPsi$ and $\bar{\bPsi}$ for respective original and interpolated pixels and interpolator $\bTheta$, we first compute joint operators $\bTheta^*$ and $\bPsi^*$, then compute output image $\y = \bTheta^* \bPsi^* \x$ for input image $\x$. } 
    \label{fig:joint_overview}
\end{figure}

In this paper, we study this joint denoising / interpolation problem from a mixed graph filtering perspective, leveraging progress in \textit{graph signal processing} (GSP)  \cite{ortega18ieee,cheung18}. 
We first prove that any (pseudo-)linear denoiser\footnote{Though \cite{chan19} proved a similar (but different) theorem for linear denoiser, our proof based on linear algebraic and convex optimization concepts is more straightforward and intuitive. See Section\;\ref{subsec:denoiser} for details.} and interpolator can be expressed as the solution filters of graph-regularized MAP estimation problems involving \textit{undirected} and \textit{directed} graph smoothness priors,  respectively (Theorem\;\ref{thm:denoiser} and \ref{thm:inter}).
These two equivalence results enable transformation of  standalone denoising and interpolation operators into graph-regularized MAP objectives;  
consequently, both tasks can be combined into a single optimization problem, leading to the derivation of jointly optimal denoising/interpolation operators.

Considering both denoising and interpolation simultaneously thus naturally leads to a \textit{mixed} graph model with both directed and undirected edges---an abstract graph framework that provides a mathematical foundation for joint denoising/interpolation optimization.
We first examine a basic joint denoising / interpolation formulation that employs two graphs: an undirected graph $\cG^u$ corresponding to denoiser $\bPsi$ inter-connecting original pixels for denoising, and a directed graph $\cG^d$ corresponding to interpolator $\bTheta$ connecting interpolated pixels to original pixels for interpolation.
We prove that this formulation leads to an optimal \textit{separable} solution that first denoises the original pixels then interpolates the denoised pixels to interpolated pixels (Corollary\;\ref{corol:separable}).

Next, we enrich the previous formulation by adding an undirected graph $\bar{\cG}^u$ corresponding to an additional denoiser $\bar{\bPsi}$ inter-connecting interpolated pixels for denoising.
We prove that this formulation leads to an optimal \textit{non-separable} solution, where new operators $\bTheta^*$ and $\bPsi^*$ are computed from original denoisers $\bPsi$ and $\bar{\bPsi}$ and interpolator $\bTheta$ (Corollary\;\ref{corol:non-separable}). 

Recovering corresponding graphs from denoisers and interpolators requires matrix inversion, which is computationally expensive.
To enable practical deployment, we develop an efficient implementation based on sparse inverse estimation \cite{Cai2011_CLIME} and iterative linear-system solvers \cite{Shewchuk1994cg}.
Experimental results show that the proposed mixed-graph framework consistently outperforms separate denoising and interpolation pipelines across a variety of denoisers, including Gaussian and bilateral filtering, and interpolation tasks, including rotation, warping, and demosaicking.

The main contributions of this paper are 
\begin{itemize} 
\item We establish graph-regularized MAP interpretations of (pseudo-)linear denoisers and interpolators using undirected and directed graph smoothness priors, respectively. 
\item We propose a mixed directed-undirected graph framework for joint image denoising and interpolation, enabling existing denoisers and interpolators to be combined within a unified optimization formulation. 
\item We characterize formulations that lead to separable and non-separable optimal solutions, showing when joint denoising/interpolation reduces to a sequential pipeline and when new jointly optimal operators emerge. 
\item We develop an efficient implementation based on sparse inverse estimation and iterative linear-system solvers. 
\item Experimental results show consistent improvements over separate denoising and interpolation pipelines across multiple denoisers, interpolators, and imaging tasks. 
\end{itemize}

%% file: related.tex
\subsection{Joint Image Denoising / Interpolation}

One well-studied joint image denoising / interpolation problem is joint demosaicking / denoising (JDD) \cite{Jin2020,Liu2020,Xing2021}. 
Given a noise-corrupted Bayer-patterned image, the joint  problem is to reconstruct full-color channels of the image by interpolating missing color components at each pixel position. 
A separate approach that performs demosaicking followed by denoising would spread the original additive noise to neighboring pixel locations during the demosaicking (interpolation) step, introducing correlated noise and making the subsequent denoising problem more challenging. 
To address the above problem, JDD schemes perform denoising and demosaicking at the same time to limit noise pollution. 
\cite{hirakawa2006joint} proposed a total least squares image denoising method. 
\cite{klatzer2016learning} tackled the JDD problem by learning a sequence of energy minimization problems. 
\cite{gharbi2016deep} proposed a deep-learning based JDD method, which interpolates missing pixels by using convolutional neural network (CNN) trained with large data, where the training data is generated by removing pixels from natural images. 
\cite{dong2018joint} proposed a Generative Adversarial Network (GAN) based approach for JDD, which uses a  discriminator network for quality assurance. 
\cite{kokkinos2019iterative} unrolled an iterative optimization algorithm into a deep network architecture for JDD, resulting in a reduction of model parameters. 
Compared with JDD methods, which are tailored specifically to demosaicking, our formulation addresses a broader class of interpolation tasks. 
Further, rather than learning a task-specific reconstruction network, we assume existing denoisers and interpolators are available and seek to combine them optimally within a unified mixed-graph optimization framework.

Our proposed optimization for joint denoising / interpolation bears some resemblance to existing \textit{plug-and-play} (PnP) ADMM frameworks \cite{venka13}, where any off-the-shelf denoiser can be utilized to solve a denoising sub-problem, as different ADMM variables are optimized in turn per iteration.
Unlike PnP methods, which embed a denoiser as a module within an iterative optimization framework, our approach derives explicit graph-regularized MAP objectives corresponding to both denoisers and interpolators. 
This enables the derivation and theoretical analysis of optimal joint denoising/interpolation operators through a unified graph-theoretic framework.

\subsection{Graph Spectral Image Restoration} 

\textit{Graph Signal Processing} (GSP)  studies computation tools to process discrete signals on graph-structured data kernels. 
In the past decade, GSP tools have been designed for various imaging tasks \cite{cheung21}: compression \cite{hu15,hu15spl}, denoising \cite{pang17,zeng19,vu21}, JPEG dequantization \cite{liu17}, deblurring \cite{bai19}, contrast enhancement \cite{liu19}, and interpolation \cite{chen21}. 
Though differing in details, the common strategy is to first construct an appropriate undirected graph capturing inter-pixel similarities, then derive an optimal graph filter by minimizing a MAP objective regularized using a graph smoothness prior, such as the \textit{graph Laplacian regularizer} (GLR) \cite{pang17}. 
The underlying graph can be computed from a statistical perspective \cite{egilmez17,bagheri21}, from a metric learning perspective \cite{hu20,yang21}, or via deep learning \cite{zeng19,vu21}. 

All the aforementioned graph construction methods employ \textit{undirected} graphs.
In contrast, we show that any full row-rank interpolator $\bTheta$ can be interpreted as a graph filter solving a MAP estimation problem regularized by a \textit{directed} graph smoothness prior called \textit{graph shift variation} (GSV) \cite{romano17}. 
A directed graph is particularly suitable for interpolation because information flows naturally from observed pixels to interpolated pixels. Unlike denoising, where neighboring pixels influence each other symmetrically, interpolated pixels should depend on original observations, while the reverse dependency is generally undesirable.
To the best of our knowledge, existing graph-based image restoration methods predominantly employ undirected graphs, whereas our framework combines directed graphs for interpolation and undirected graphs for denoising within a unified optimization formulation.

Compared to the conference version of this paper \cite{Viswarupan2024}, we extend our work in the following non-trivial ways:
i) we examine the general non-separable joint denoising / interpolation scenario with two denoisers for both the original pixels and interpolated pixels; ii) we propose a fast filter implementation based on sparse inverse matrix estimation \cite{Cai2011_CLIME} and conjugate gradient \cite{Shewchuk1994cg} for the non-separable case; and iii) we generalize our analysis to rectangular interpolator matrices via multiple interpolators decomposition.

%% file: prelim.tex
\subsection{GSP Definitions}

We review the graph signal processing concepts used throughout the paper \cite{ortega18ieee}.
A graph $\mathcal{G}=(\mathcal{V}, \mathcal{E})$ consists of a node set $\mathcal{V}$ of size $N$ and an edge set $\mathcal{E}$ specified by $(i,j,w_{i,j})$, where $i,j \in \mathcal{V}$ and $w_{i,j} \in \mathbb{R}$ is a scalar weight of an edge $(i,j) \in \cE$ encoding the (dis)similarity between samples at nodes $i$ and $j$. 
We define an \textit{adjacency matrix} $\A \in \mathbb{R}^{N \times N}$, where $A_{i,j} = w_{i,j}$ if $(i,j) \in \cE$, and $A_{i,j}=0$ otherwise. 
We consider both \textit{undirected} and \textit{directed} graphs.
An undirected graph $\cG$ means $A_{i,j} = A_{j,i}, \forall i,j \in \cV$, and $\A$ is symmetric. 

We next define a diagonal \textit{degree matrix} $\mathbf{D} \in \mathbb{R}^{N \times N}$, where $D_{i,i}=\sum_{j} A_{i,j}$. 
Given $\A$ and $\D$, we define a \textit{combinatorial graph Laplacian matrix} as $\mathbf{L} \triangleq \mathbf{D}-\A$.
By assigning sample value $x_i \in \mathbb{R}$ to node $i$, the ensemble $\x = [x_1 \ldots x_N]^{\top} \in \mathbb{R}^N$ is interpreted as a \textit{graph signal} on graph $\mathcal{G}$.

A \textit{symmetric normalized graph Laplacian} $\L_n \triangleq \D^{-1/2} \L \D^{-1/2}$ has eigenvalues in range $[0,2]$.
A \textit{random walk graph Laplacian matrix} $\L_{r} \triangleq \I - \D^{-1} \A = \D^{-1} \L$, a similarity transform of $\L_n$ (\ie, $\L_{r} = \D^{-1/2} \L_n \D^{1/2}$), shares the same eigenvalues as $\L_n$ but is asymmetric.

If graph $\cG$ has self-loops, \ie, $\exists i$ such that $w_{i,i} > 0$, then the \textit{generalized graph Laplacian matrix} is typically used instead, defined as $\cL \triangleq \D - \A + \mathrm{diag}(\A)$, where $\mathrm{diag}(\A)$ returns a matrix with only the diagonal entries of $\A$.

If $\cG$ is a \textit{signed} graph with positive and negative edges, then often a \textit{signed graph Laplacian matrix} $\L^s \triangleq \D^s - \A$ is used, where degree for node $i$ is $D^s_{i,i} = \sum_j |A_{i,j}|$ \cite{Kunegis2010}.
This definition ensures graph Laplacian $\L^s$ is \textit{positive semi-definite} (PSD).
In this paper, we use the generalized graph Laplacian for both positive and signed undirected graphs. 
Under the conditions established later in Theorem~\ref{thm:denoiser}, the resulting Laplacian is guaranteed to be PSD.

The \textit{graph spectrum} of an undirected graph $\cG$ is the eigen-pairs $\{(\lambda_k, \v_k)\}_{k=1}^N$ of Laplacian $\L$, where eigenvalue $\lambda_k$ is the $k$-th frequency, and eigenvector $\v_k$ is the $k$-th graph Fourier mode. 
The collection of eigenvectors $\{\v_k\}_{k=1}^N$ compose the \textit{graph Fourier transform} (GFT) $\V^{\top}$, where $\L = \V \mathrm{diag}(\{\lambda_k\}) \V^{\top}$. $\tilde{\x} = \V^{\top} \x$ are the GFT coefficients of signal $\x$ \cite{ortega18ieee}.

\subsection{Graph Smoothness Priors}

There exist several graph smoothness priors in the GSP literature; each assumes signal $\x$ is smooth w.r.t. the underlying graph $\cG$ but is expressed in slightly different mathematical terms. 
The most common is the \textit{graph Laplacian regularizer} (GLR) \cite{pang17}:
\begin{align}
\x^\top \L \x = \sum_{(i,j)\in\cE} w_{i,j}(x_i-x_j)^2 = \sum_k \lambda_k \tilde{x}_k^2  
\label{eq:GLR}
\end{align}
where $\L$ is the combinatorial graph Laplacian for graph $\cG$, $\lambda_k$ is the $k$-th eigenvalue of $\L$, and $\tilde{x}_k$ is the corresponding $k$-th GFT coefficient for signal $\x$ upon computing $\tilde{\x} = \V^{\top} \x$. 
If edge weights are non-negative, \ie, $w_{i,j} \geq 0, \;\forall (i,j) \in \cE$, then $\L$ is provably PSD and $\x^\top \L \x \geq 0, \forall \x \in \mathbb{R}^N$ \cite{cheung18}. 
From \eqref{eq:GLR}, in the nodal domain a small GLR means a connected node-pair $(i,j)$ with large edge weight $w_{i,j}$ should have similar sample values $x_i$ and $x_j$.
On the other hand, in the graph frequency domain a small GLR means signal energies $\tilde{x}_k^2$ should reside mostly in low frequencies $\lambda_k$, \ie, signal $\x$ should be low-pass. 

Another common graph smoothness prior is the \textit{graph shift variation}\footnote{A similar definition from \cite{chen15} is a graph variation term $\|\x - \frac{1}{\lambda_{\max}} \A \x\|^2_2$, where $\A$ is a graph adjacency matrix interpreted as a shift operator. We adopt the definition in \cite{romano17} as GSV instead, since it does not require the computation of the largest eigenvalue $\lambda_{\max}$ beforehand. } (GSV) \cite{romano17}.
First, define the row-stochastic adjacency matrix as $\A_{r} = \D^{-1} \A$. 
We then write GSV as
\begin{align}
\| \x - \A_r \x\|^2_2 &= \|(\I - \A_r)\x\|^2_2 = \| \L_r \x \|^2_2 
\label{eq:GSV} \\
&= \sum_k \lambda_k \tilde{x}_k^2
\nonumber 
\end{align}
where $\L_r$ is the random-walk graph Laplacian, and $\lambda_k$ are the eigenvalues of $\L_r^\top \L_r$. 
$\tilde{x}_k$ is the $k$-coefficient in the expansion $\V^{\top} \x$, where $\V$ are the orthonormal eigenvectors of symmetric $\L_r^\top \L_r$ in columns, \ie, $\L_r^\top \L_r = \V \mathrm{diag}(\{\lambda_k\}) \V^{\top}$. 
GSV \eqref{eq:GSV} can be interpreted as the $\ell_2$-norm difference between signal $\x$ and its shifted version $\A_r \x$, where the row-stochastic adjacency matrix $\A_r$ is the \textit{graph shift operator} (GSO) \cite{ortega18ieee}. 
Like GLR in \eqref{eq:GLR}, a small GSV means that $\x$ has its energies residing mostly in low graph frequencies $\lambda_k$, where the frequencies are defined by the spectrum of $\L_r^\top\L_r$. 
GSV can be alternatively written as $\x^{\top} \L_r^{\top} \L_r \x$, called \textit{left eigenvectors of the random walk graph Laplacian} (LERaG) in \cite{liu17}. 
It was proposed for JPEG image dequantization as a normalized signal prior that evaluates to zero for a constant signal.

In contrast to GLR \eqref{eq:GLR}, one important characteristic of GSV \eqref{eq:GSV} is that it is well defined even if the graph $\cG$ is directed. 
This property is crucial for our interpolation framework, where information flows directionally from observed pixels to interpolated pixels.

There exist other graph smoothness priors in the literature, such as \textit{graph total variation} (GTV) \cite{elmoataz08,couprie13,bai19}, \textit{gradient graph Laplacian regularizer} (GGLR) \cite{chen21,chen24}, and more recently, \textit{non-convex graph total variation} (NC-GTV) \cite{Wei2025}. 
In this paper, we focus on GLR and GSV, not only because they are commonly used, but because they are in convenient quadratic form, so that when they are used as signal priors along with quadratic fidelity terms, the corresponding solutions are linear filters that can be mapped to linear denoisers and interpolators, to be discussed in detail next.

%% file: graphInterpolate.tex
\subsection{Denoiser: Solution of Graph-Regularized MAP Problem}
\label{subsec:denoiser}

We first establish a theorem (different from one in \cite{chan19}) that relates a denoising matrix $\bPsi \in \mathbb{R}^{N \times N}$ to a graph-regularized \textit{maximum a posteriori} (MAP) optimization problem.
While \cite{chan19} utilized a proximal map property \cite{parikh13} to substantiate the relationship, our derivation is purely linear algebraic, providing a simpler and more direct interpretation.  

Consider a linear denoising operation written as
\begin{align}
\x = \bPsi \y 
\label{eq:denoiser}
\end{align}
where $\x, \y \in \mathbb{R}^N$ are the output and input signals of the denoiser, respectively. 
Theorem 1 in \cite{chan19} assumes that $\bPsi$ is symmetric, doubly stochastic\footnote{Entries of each row / column of a doubly stochastic matrix are non-negative \textit{and} sum to one. 
The overview paper \cite{milanfar13} on image filtering also assumes that the filtering matrix is doubly stochastic.}, and invertible.
The doubly stochastic property implies i) entries in $\bPsi$ are non-negative, and ii) \textit{non-expansiveness}, meaning $\|\bPsi \x\|^2_2 \leq \|\x\|^2_2, \forall \x$, which in turn means that eigenvalues $\{\lambda_k\}$ of $\bPsi$ satisfies $|\lambda_k| \leq 1, \forall k$. 
In contrast, we assume that $\bPsi$ is non-expansive, symmetric, and \textit{positive definite} (PD), \ie, $0 < \lambda_k \leq 1, \forall k$.  
Thus, unlike \cite{chan19}, we require eigenvalues $\lambda_k$'s to be strictly positive, but do not require $\bPsi \1 = \1$. 
Further, we impose no requirement for entries in $\bPsi$ to be non-negative.

A \textit{signal-dependent} denoiser $\bPsi(\x)$, such as the well-known \textit{bilateral filter} (BF) \cite{tomasi98}, that depends on the target signal $\x$ to first determine filter weights, is not linear in general.
However, one can nonetheless consider it \textit{pseudo-linear} \cite{chan19}: 
for each iteration $t$ in an iterative filtering procedure, we first compute a linear filter $\bPsi(\x_{t})$ using solution $\x_{t}$ obtained from the previous iteration $t-1$, then the filtering $\bPsi \x_{t}$ is a linear operation \cite{milanfar13}---a matrix-vector multiplication.

Consider next the following standard MAP optimization problem for denoising input $\y$, using GLR \eqref{eq:GLR} \cite{pang17} as the signal prior:
\begin{align}
\min_\x \|\y - \x\|^2_2 + \mu \, \x^{\top} \L \x 
\label{eq:MAP_denoise}
\end{align}
where $\mu > 0$ is a weight parameter, and $\L$ is a combinatorial graph Laplacian matrix for an undirected graph $\cG$ (hence $\L$ is symmetric).
Assuming $\L$ is PSD, \eqref{eq:MAP_denoise} is an unconstrained and convex \textit{quadratic programming} (QP) problem with solution
\begin{align}
\x^* = \left(\I_N + \mu \L \right)^{-1} \y 
\label{eq:MAP_sol_denoise}
\end{align}
where $\I_N$ is an $N \times N$ identity matrix. 
Note that coefficient matrix $\I_N + \mu \L$ is provably \textit{positive definite} (PD) via Weyl's Inequality \cite{HornJohnson2012} and thus invertible. 

We now state a theorem that connects denoiser $\bPsi$ \eqref{eq:denoiser} and the MAP problem \eqref{eq:MAP_denoise} as follows:
\begin{theorem}
Denoiser $\bPsi$ \eqref{eq:denoiser} is the solution filter for the MAP problem \eqref{eq:MAP_denoise} if $\L = \mu^{-1} (\bPsi^{-1} - \I_N)$, assuming matrix $\bPsi$ is non-expansive, symmetric, and PD. 
\label{thm:denoiser}
\end{theorem}

\begin{proof}
Symmetry means $\bPsi$ has real eigenvalues.
Non-expansiveness and PDness mean $\bPsi$ is invertible with positive eigenvalues $0 < \lambda_k \leq 1, \forall k$.
Thus, $\bPsi^{-1}$ exists and has positive eigenvalues $\{\lambda_k^{-1}\}$. 
Condition $\lambda_k \leq 1, \forall k$ means that $1 \leq \lambda_k^{-1}, \forall k$.
Thus, the eigenvalues of $\bPsi^{-1} - \I_N$ are $\lambda_k^{-1}-1 \geq 0, \forall k$.
This implies $\L = \mu^{-1} (\bPsi^{-1} - \I_N)$ is PSD, and thus quadratic objective \eqref{eq:MAP_denoise} is convex.
Taking derivative w.r.t. $\x$ and setting it to zero, MAP optimization \eqref{eq:MAP_denoise} has \eqref{eq:MAP_sol_denoise} as solution. 
Inserting $\L = \mu^{-1} (\bPsi^{-1} - \I_N)$ into \eqref{eq:MAP_sol_denoise}, we get $\x^* = \bPsi \y$, and thus $\bPsi$ is the resulting solution filter.
\end{proof}

\noindent 
\textbf{Remarks}: 
In general, PSD matrix $\L = \mu^{-1} (\bPsi^{-1} - \I_N)$ corresponding to non-expansive, symmetric and PD $\bPsi$ is a generalized graph Laplacian of an undirected graph with positive / negative edges and self-loops.
This class of denoisers $\bPsi$ includes a broad family of pseudo-linear denoisers satisfying these conditions, \ie, symmetric PD operators with spectral norm at most $1$, which commonly arise in kernel-based and diffusion-based filtering \cite{milanfar13,tomasi98,buades05,Perona1990}.
Theorem\;\ref{thm:denoiser} implies that an undirected graph filter---solution to undirected-graph-regularized MAP problem \eqref{eq:MAP_denoise}---is as expressive as a linear denoiser $\bPsi$ under specified assumptions. 

One benefit of Theorem\;\ref{thm:denoiser} is \textit{interpretability}: 
any linear denoiser $\bPsi$ satisfying the aforementioned requirements can be interpreted as a \textit{graph filter} associated with an (typically sparse) \textit{undirected} graph $\cG^u$, specified by $\L = \mu^{-1}(\bPsi^{-1} - \I_N)$, where the symmetry of $\L$ induces an undirected topology. 
In fact, \textit{bilateral filtering} (BF) \cite{tomasi98} has been interpreted as a graph filter in the GSP literature \cite{gadde13}; Theorem\;\ref{thm:denoiser} generalizes this connection to a broader class of pseudo-linear denoisers.

\subsection{Interpolator: Solution of Graph-Regularized MAP Problem}

We next investigate a linear interpolator $\bTheta \in \mathbb{R}^{N \times M}$ that interpolates $N$ new pixels from $M$ original pixels $\y \in \mathbb{R}^M$.  
Mathematically, we write
\begin{align}
\x = \left[ \begin{array}{c}
\x_M \\
\x_N 
\end{array} \right] 
= \left[ \begin{array}{c}
\I_{M} \\
\bTheta
\end{array} \right] \y 
\label{eq:interpolator}
\end{align}
where $\x \in \mathbb{R}^{M+N}$ is the length-$(M+N)$ target signal that retains the original $M$ pixels, \ie, $\x_M = \y$.  

We define a MAP optimization objective for interpolation, similar to \eqref{eq:MAP_denoise} for denoising.
Denote by $\H = [\I_{M} ~~ \0_{M,N}]$ a $M \times (M+N)$ \textit{sampling matrix} that selects $M$ original pixels, $\x_M$, from signal $\x$, where $\0_{M,N}$ is a $M \times N$ matrix of zeros.  
Denote by $\A \in \mathbb{R}^{(M+N) \times (M+N)}$ an \textit{asymmetric} adjacency matrix specifying \textit{directional} edges in a directed graph $\cG^d$ for signal $\x$.
Specifically, $\A$ describes edges \textit{only} from the $N$ new pixels to the $M$ original pixels, \ie, 
\begin{align}
\A = \left[ \begin{array}{cc}
\0_{M,M} & \A_{M,N} \\
\0_{N,M} & \0_{N,N}
\end{array} \right] .
\label{eq:A_directed}
\end{align}
$\A \x = [\A_{M,N} \x_N; \0_N]$ thus maps $N$ new pixels $\x_N$ back to the $M$ original pixels via $\A_{M,N}$. 

We now write a MAP optimization objective using GSV \eqref{eq:GSV} as the signal prior:
\begin{align}
\min_{\x} \|\y - \H \x\|^2_2 + \gamma \|\H (\x - \A \x)\|^2_2 
\label{eq:MAP_inter}
\end{align}
where $\gamma > 0$ is a weight parameter. 
GSV $\|\H (\x - \A \x) \|^2_2$ states that a smooth graph signal $\x$ should be similar to its shifted version $\A \x$, but we evaluate only the $M$ original pixels in the objective. 
Note that \eqref{eq:MAP_inter} is convex for any definition of $\A$. 

We state formally a theorem to connect interpolator in \eqref{eq:interpolator} and the MAP problem \eqref{eq:MAP_inter}. 
\begin{theorem}
The interpolator $[\I_M; \bTheta]$ \eqref{eq:interpolator} is the solution filter to the MAP problem \eqref{eq:MAP_inter} if $M=N$, $\bTheta$ is invertible, and $\A_{M,N} = \bTheta^{-1}$.
\label{thm:inter}
\end{theorem}

\begin{proof}
we rewrite $\H(\x-\A\x) = \H(\I-\A)\x$.
Given that \eqref{eq:MAP_inter} is convex for any $\A$, we take the derivative w.r.t. $\x$ and set it to $0$, resulting in

\vspace{-0.1in}
\begin{small}
\begin{align}
\left(\H^{\top} \H + \gamma (\I-\A)^\top \H^\top \H (\I-\A) \right) \x = \H^{\top} \y .
\label{eq:MAP_inter_linSys}
\end{align}
\end{small}\noindent
Given the definitions of $\H$ and $\A$, we rewrite \eqref{eq:MAP_inter_linSys} as
\begin{align}
\underbrace{\left[ \begin{array}{cc}
(1 + \gamma) \I_M  & -\gamma \A_{M,N} \\
- \gamma \A_{M,N}^\top & \gamma \A^2_{N,N}
\end{array} \right]}_{\C} \x = \H^{\top} \y
\label{eq:MAP_inter_linSys2}
\end{align}
where $\A^2_{N,N} \triangleq \A_{M,N}^\top \A_{M,N}$.
Note that though $M=N$ by assumption, $\A_{M,N}$ is asymmetric in general, and thus $\A_{M,N}^\top \neq \A_{M,N}$.

Using a matrix inversion formula \cite{bernstein2009matrix},
\begin{footnotesize}
\begin{align}
\left[ \begin{array}{cc}
\dot{\A} & \dot{\B} \\
\dot{\C} & \dot{\D} 
\end{array} \right]^{-1}
&= \left[ \begin{array}{cc}
\dot{\P} & -\dot{\P} \dot{\B} \dot{\D}^{-1} \\
-\dot{\D}^{-1} \dot{\C} \dot{\P} & \dot{\D}^{-1} + \dot{\D}^{-1} \dot{\C} \dot{\P} \dot{\B} \dot{\D}^{-1}
\end{array} \right] 
\label{eq:inverse}
\end{align}
\end{footnotesize}\noindent 
where $\dot{\P} = (\dot{\A} - \dot{\B} \dot{\D}^{-1} \dot{\C})^{-1}$, we first compute $\dot{\P}$ for coefficient matrix $\C$ in \eqref{eq:MAP_inter_linSys2} as
\begin{align}
\dot{\P} &= \left( (1+\gamma) \I_M - (-\gamma \A_{M,N}) (\gamma \A_{N,N}^2)^{-1} (-\gamma \A_{M,N}^\top) \right)^{-1} 
\nonumber \\
&= \left( (1+\gamma) \I_M - \gamma \A_{M,N} (\A_{M,N}^\top \A_{M,N})^{-1} \A_{M,N}^\top \right)^{-1}
\nonumber \\
&\stackrel{(a)}{=} \left( (1+\gamma) \I_M - \gamma \A_{M,N} \A_{M,N}^{-1} (\A_{M,N}^\top)^{-1} \A_{M,N}^\top \right)^{-1}
\nonumber \\
&= \left( (1+\gamma) \I_M - \gamma \I_M \right)^{-1} = \I_M ,
\end{align}
where in $(a)$ we apply the assumption that $\A_{M,N} = \bTheta^{-1}$ and thus invertible.

We can now write $\x$ from \eqref{eq:MAP_inter_linSys2} as

\vspace{-0.1in}
\begin{small}
\begin{align}
\x &= \C^{-1} \H^\top \y = \left[ \begin{array}{c}
\I_M \\
-(\gamma \A_{N,N}^2)^{-1} (-\gamma \A_{M,N}^\top) \I_M 
\end{array} \right] \y \nonumber \\
&= \left[ \begin{array}{c}
 \I_M \\ \A_{M,N}^{-1} (\A_{M,N}^\top)^{-1} \A_{M,N}^\top
\end{array} \right] \y 
= \left[ \begin{array}{c}
\I_M \\
\A^{-1}_{M,N} 
\end{array} \right] \y .
\label{eq:Q}
\end{align}
\end{small}\noindent
Given $\A_{M,N}^{-1} = \bTheta$ by assumption, we conclude that
\begin{align}
\x = \left[ \begin{array}{c}
\I_M \\
\bTheta \end{array} \right] \y .
\end{align}
Hence, $[\I_M; \bTheta]$ is the solution filter to \eqref{eq:MAP_inter}. 
\end{proof}

\noindent
\textbf{Remarks}: 
Though Theorem\;\ref{thm:inter} requires $M=N$, we retain the notation $\A_{M,N}$ to stress that the interpolator is, in general, asymmetric, and thus naturally associated with a directed graph.
Theorem\;\ref{thm:inter} implies that, under the stated conditions, a directed graph filter---\ie, the solution to the MAP problem \eqref{eq:MAP_inter} regularized by a \textit{directed} graph—is as expressive as a linear interpolator $\bTheta$.
The key requirements are that $M=N$ and $\bTheta$ is invertible (full rank), corresponding to a non-degenerate interpolation operator with linearly independent rows (or columns).
Such non-degenerate (well-posed) interpolation operators arise in a broad class of practical interpolators, including kernel-based interpolation with distinct samples \cite{Micchelli1986}, local polynomial and regression-based methods with full-rank design matrices \cite{FanGijbels1996}, spline interpolation \cite{deBoor1978}, and moving least-squares methods for scattered data \cite{Levin1998}.

Intuitively, using \textit{bidirectional} edges $(i,j)$ in an \textit{undirected} graph for denoising makes sense; 
the uncertainty in observed noisy pixels $i$ and $j$ means that their reconstructions depend on each other.
In contrast, using \textit{directional} edge $(i,j)$ in a \textit{directed} graph for interpolation is reasonable; interpolated pixel $j$ depends on original pixel $i$ but not vice versa.

Note that denoiser $\bPsi$ in \eqref{eq:denoiser} is required to be non-expansive, symmetric, and PD, so that its real eigenvalues obey $0 < \lambda_k \leq 1, \forall k$.
In contrast, interpolator $\bTheta$ in \eqref{eq:interpolator} has no restrictions beyond being square and full rank (no zero eigenvalues).
This is because optimization \eqref{eq:MAP_inter} is convex for any definition of $\A$.
Nonetheless, the graph specification for interpolation---adjacency matrix $\A_{M,N}$ in \eqref{eq:A_directed}---involves a matrix inverse of $\bTheta$ in Theorem\;\ref{thm:inter}.
In this sense, Theorem\;\ref{thm:inter} and \ref{thm:denoiser} are similar.

%% file: JDI.tex
Having developed Theorem\;\ref{thm:denoiser} and Theorem\;\ref{thm:inter} to relate linear denoiser $\bPsi$ and linear interpolator $\bTheta$ to graph-regularized MAP problems \eqref{eq:MAP_denoise} and \eqref{eq:MAP_inter}, respectively, we analyze different joint denoising / interpolation formulations in this section.

\subsection{Joint Formulation with Separable Solution}

\begin{figure}[t]
    \vspace{-1em}
    \centering
    \begin{minipage}{0.48\linewidth}
        \centering
        \includegraphics[width=\linewidth]{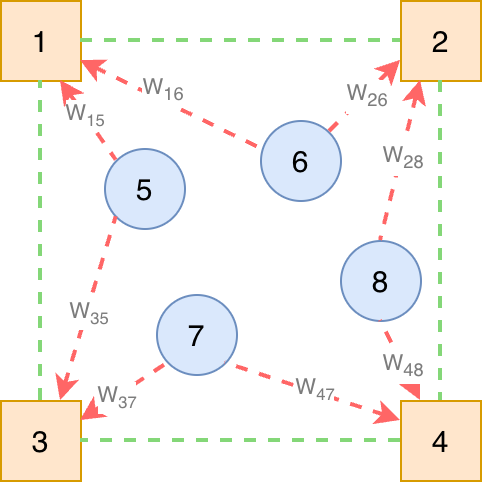}
        \subcaption{}
    \end{minipage}
    \hfill
    \begin{minipage}{0.48\linewidth}
        \centering
        \includegraphics[width=\linewidth]{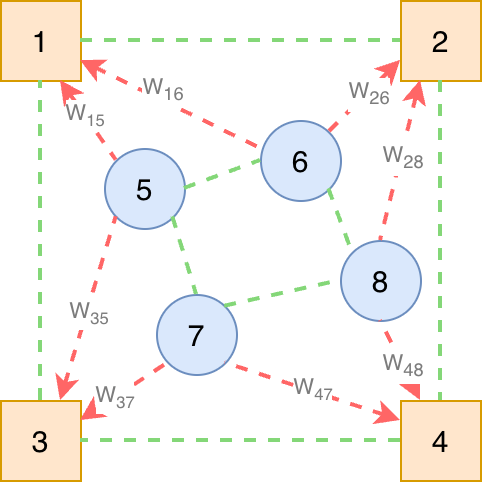}
        \subcaption{}
    \end{minipage}
    \vspace{-0.05in}
    \caption{Illustrations of mixed graphs for joint interpolation / denoising. Pixels 5,6,7,8 are interpolated using pixels 1,2,3,4. Directed edges for interpolation are shown as red dashed arrows, while undirected edges for denoising are shown as green dashed lines. (a) corresponds to corollary\;\ref{corol:separable} where input pixels are denoised before interpolation. (b) corresponds to corollary\;\ref{corol:non-separable} where input pixels are denoised before interpolation and interpolated pixels are denoised after interpolation.}
    \label{fig:image2}
\end{figure}

Denote by $\A \in \mathbb{R}^{(M+N) \times (M+N)}$ in the form \eqref{eq:A_directed} an adjacency matrix of a \textit{directed} graph $\cG^d$ connecting $N$ new pixels to $M$ original pixels, corresponding to an interpolator $\bTheta$. 
Further, denote by $\L \in \mathbb{R}^{M \times M}$ a graph Laplacian matrix for an \textit{undirected} graph $\cG^u$ inter-connecting the $M$ noisy original pixels, corresponding to a denoiser $\bPsi$. 
See Fig.\;\ref{fig:image2}(a) for an illustrative example, where new pixels $5,6,7,8$ are interpolated using original pixels $1,2,3,4$.
Here, interpolator $\bTheta$ is visualized using \textit{directional} edges of directed graph $\cG^d$ from new pixels $5,6,7,8$ to original pixels $1,2,3,4$, while denoiser $\bPsi$ is visualized using \textit{bidirectional} edges of undirected graph $\cG^u$ connecting original pixels $1,2,3,4$.

One formulation for joint denoising / interpolation is to simply combine terms in MAP objectives \eqref{eq:MAP_denoise} and \eqref{eq:MAP_inter} as

\vspace{-0.05in}
\begin{small}
\begin{align}
\min_{\x} \|\y - \H \x\|^2_2 + \gamma \| \H(\x-\A\x)\|^2_2 + \mu (\H \x)^{\top} \L (\H \x) .
\label{eq:MAP_mixed}
\end{align}
\end{small}\noindent
In words, \eqref{eq:MAP_mixed} states that sought signal $\x$ should be smooth w.r.t. \textit{two} graphs $\cG^d$ and $\cG^u$: i) $\x$ should be similar to its shifted version $\A \x$ where $\A$ is an adjacency matrix for a \textit{directed} graph $\cG^d$, and ii) original pixels $\H \x$ should be smooth w.r.t. to an \textit{undirected} graph $\cG^u$ defined by $\L$. 
This formulation follows our intuition developed in Section\;\ref{sec:graphInterpolate}: 
use \textit{directional} edges to connect interpolated pixels to original ones, and use \textit{bidirectional} edges to inter-connect noisy pixels.

It turns out that the optimal solution to the posed MAP problem \eqref{eq:MAP_mixed} takes a particular form.
\begin{corollary}
\eqref{eq:MAP_mixed} has a separable optimal solution. 
\label{corol:separable}
\end{corollary}

\begin{proof}
Optimization \eqref{eq:MAP_mixed} is an unconstrained convex QP problem.
Taking the derivative w.r.t. $\x$ and setting it to $0$, we get the following linear system:

\vspace{-0.05in}
\begin{footnotesize}
\begin{align}
\left( \H^{\top} \H + \gamma \left( (\I-\A)^\top \H^{\top} \H (\I - \A) \right) + \mu \H^{\top} \L \H \right) \x^* = \H^{\top} \y .
\label{eq:MAP_mixed_linSys}
\end{align}
\end{footnotesize}\noindent
Denote by $\C$ the coefficient matrix on the left-hand side.
Given $\H = [\I_M ~ \0_{M,N}]$, $\H^\top \L \H$ has nonzero term $\L$ only in the upper-left sub-matrix block. 
Hence, $\C$ differs from \eqref{eq:MAP_inter_linSys2} only in the additional $\mu \L$ in the upper-left block, \ie, 
\begin{align}
\C = \left[ \begin{array}{cc}
(\gamma+1) \I_M + \mu \L & -\gamma \A_{M,N} \\
-\gamma \A_{M,N}^\top & \gamma \A^2_{N,N}
\end{array} \right] .
\label{eq:coeff2_mixed}
\end{align}

Using the matrix inverse formula \eqref{eq:inverse}, we can first write block $\dot{\P} = (\dot{\A} - \dot{\B} \dot{\D}^{-1} \dot{\C})^{-1}$ as:

\vspace{-0.1in}
\begin{small}
\begin{align}
\dot{\P} &= \left( (\gamma+1)\I_M + \mu \L - (-\gamma \A_{M,N}) (\gamma \A_{N,N}^2)^{-1} (-\gamma \A_{M,N}^\top) \right)^{-1}  
\nonumber \\
&= \left( (\gamma+1)\I_M + \mu \L - \gamma \I_M \right)^{-1} = (\I_M + \mu \L)^{-1} .
\end{align}
\end{small}

Thus, the solution $\x^*$ for optimization \eqref{eq:MAP_mixed_linSys} is
\begin{align}
\x^* &= \C^{-1} \H^\top \y 
\nonumber \\
&= 
\left[ \begin{array}{c}
(\I_M + \mu \L)^{-1} \\
- (\gamma \A_{N,N}^2)^{-1} (-\gamma \A_{M,N}^\top) (\I_M + \mu \L)^{-1}
\end{array} \right] \y
\nonumber \\
&= \left[ \begin{array}{c}
(\I_M + \mu \L)^{-1} \\
\A_{M,N}^{-1} (\I_M + \mu \L)^{-1}
\end{array} \right] \y = 
\left[ \begin{array}{c}
\bPsi \\
\bTheta \, \bPsi
\end{array} \right] \y
\label{eq:MAP_mixed_linSys2}
\end{align}
where $\bPsi = \left(\I_M + \mu \L \right)^{-1}$ is the denoiser, and $\bTheta = \A_{M,N}^{-1}$ is the interpolator. 
We see that solution \eqref{eq:MAP_mixed_linSys2} to optimization \eqref{eq:MAP_mixed} is \textit{separable}: input $\y$ first undergoes denoising via original denoiser $\bPsi$, then subsequently interpolation via original interpolator $\bTheta$.
\end{proof}

\vspace{0.05in}
\noindent
\textbf{Remarks}: 
Intuitively, one can see graphically why problem \eqref{eq:MAP_mixed} has a separable solution \eqref{eq:MAP_mixed_linSys2}: the $M$ original pixels are inter-connected only via bidirectional edges (for denoising), while the $M$ original pixels are connected as end-points of directional edges (for interpolation). 
\textit{The $N$ new pixels depend entirely on the original $M$ pixels, and thus play no role in the denoising of original pixels.} 
To have an optimal non-separable denoising / interpolation solution, $N$ new pixels must be connected by bidirectional edges also.

\subsection{Joint Formulation with Non-separable Solution} 

We consider such a scenario next, where we introduce an additional GLR smoothness prior \cite{pang17} for the $N$ interpolated pixels to the formulation in \eqref{eq:MAP_mixed}, resulting in
\begin{align}
\min_{\x} & \|\y - \H \x\|^2_2 + \gamma \| \H (\x - \A \x)\|^2_2 + \mu (\H \x)^\top \L (\H \x)   
\nonumber \\
& + \kappa (\G \x)^\top \bar{\L} (\G \x)
\label{eq:MAP_mixed2}
\end{align}
where $\G = [\0_{N,M} \; \I_{N}]$ is the complementary sampling matrix of $\H$ that selects only the $N$ interpolated pixels from signal $\x$, and $\bar{\L} \in \mathbb{R}^{N \times N}$ denotes a graph Laplacian matrix for an undirected graph $\bar{\cG}^u$ connecting the interpolated pixels. 
See Fig.\;\ref{fig:image2}(b) for an illustrative example that is modified from the example in Fig.\;\ref{fig:image2}(a), where an additional denoiser $\bar{\bPsi}$ for the four interpolated pixels $5,6,7,8$ is added.
$\bar{\bPsi}$ is visualized using \textit{bidirectional} edges of undirected graph $\bar{\cG}^u$ connecting interpolated pixels $5,6,7,8$.

\begin{corollary}
\eqref{eq:MAP_mixed2} has a non-separable optimal solution.
\label{corol:non-separable}
\end{corollary}

\begin{proof}
Like \eqref{eq:MAP_mixed},  \eqref{eq:MAP_mixed2} remains convex, quadratic and differentiable. 
The optimal solution $\x^*$ can be computed via a system of linear equations,

\vspace{-0.05in}
\begin{small}
\begin{align}
\left( \H^{\top} \H + \gamma \left( (\I - \A)^\top \H^\top \H (\I - \A) \right) + \cL \right) \x^* = \H^{\top} \y ,
\label{eq:MAP_mixed2_linSys}
\end{align}
\end{small}\noindent
where $\cL = \mu \H^\top \L \H + \kappa \G^\top \bar{\L} \G$ combines the two GLR priors corresponding to undirected graphs $\cG^u$ and $\bar{\cG}^u$ and is block-diagonal, \ie,
\begin{align}
\cL = \left[ \begin{array}{cc}
\mu \L & \0_{M,N} \\
\0_{N,M} & \kappa \bar{\L}
\end{array} \right] .
\end{align}
A similar derivation shows that coefficient matrix $\C$ changes from \eqref{eq:coeff2_mixed} to
\begin{align}
\C = \left[ \begin{array}{cc}
(\gamma+1) \I_M + \mu \L & -\gamma \A_{M,N} \\
-\gamma \A_{M,N}^\top & \gamma \A^2_{N,N} + \kappa \bar{\L}
\end{array} \right] .
\label{eq:coeff2_mixed_mod}
\end{align}

Using again the matrix inverse formula \eqref{eq:inverse}, we first write block $\dot{\P} = (\dot{\A} - \dot{\B} \dot{\D}^{-1} \dot{\C})^{-1}$ as

\vspace{-0.1in}
\begin{footnotesize}
\begin{align}
\dot{\P} &= \left( (\gamma+1) \I_M + \mu \L - (-\gamma \A_{M,N}) (\gamma \A_{N,N}^2 + \kappa \bar{\L} )^{-1} (-\gamma \A_{M,N}^\top)
\right)^{-1} 
\nonumber \\
&= \left( (\gamma+1) \I_M + \mu \L - \gamma^2 \A_{M,N} (\gamma \A_{N,N}^2 + \kappa \bar{\L} )^{-1} \A_{M,N}^\top
\right)^{-1} .
\end{align}
\end{footnotesize}\noindent 
The solution for $\x^*$ in \eqref{eq:MAP_mixed2_linSys} is

\vspace{-0.1in}
\begin{small}
\begin{align}
\x^* &= \C^{-1} \H^\top \y 
\nonumber \\
&= \left[ \begin{array}{c}
\dot{\P} \\
-(\gamma \A_{N,N}^2 + \kappa \bar{\L})^{-1} (-\gamma \A_{M,N}^\top) \dot{\P}
\end{array} \right] \y
\nonumber \\
&= \left[ \begin{array}{c}
\dot{\P} \\
\gamma (\gamma \A_{N,N}^2 + \kappa \bar{\L})^{-1} \A_{M,N}^\top \dot{\P}
\end{array} \right] \y =
\left[ \begin{array}{c}
\bPsi^* \\
\bTheta^* \, \bPsi^*
\end{array} \right] \y
\label{eq:MAP_mixed2_linSys2}
\end{align}
\end{small}\noindent
where $\bPsi^* = \dot{\P}$ and $\bTheta^* = \gamma (\gamma \A_{N,N}^2 + \kappa \bar{\L})^{-1} \A_{M,N}^\top$ are jointly optimal denoising and interpolation operators induced by the mixed-graph optimization.
Since $\bTheta^* \bPsi^* \neq \bar{\bPsi} \bTheta \bPsi$, the solution to \eqref{eq:MAP_mixed2} is generally non-separable\footnote{Though the solution $\bTheta^* \bPsi^*$ to \eqref{eq:MAP_mixed2} is a sequence of two separate matrix operations, each matrix is a non-separable function of input matrices $\{\A, \L, \bar{\L}\}$ or $\{\bTheta, \bPsi, \bar{\bPsi}\}$.}.
\end{proof}

\vspace{0.05in}
\noindent
\textbf{Remarks}: 
In this setting, the interpolator is no longer an independent downstream operation; it now influences the denoising operation.
If $\kappa = 0$ (\ie, no second denoiser $\bar{\bPsi}$ for $N$ interpolated pixels), then $\bPsi^* = \bPsi$, $\bTheta^* = \bTheta$, \eqref{eq:MAP_mixed2_linSys2} defaults to \eqref{eq:MAP_mixed_linSys2}, and the solution becomes separable. 
On the other hand, if $\mu = 0$ (\ie, no first denoiser $\bPsi$ for $M$ original pixels\footnote{This case was discussed in the conference version \cite{Viswarupan2024} of this paper.}), then

\vspace{-0.05in}
\begin{small}
\begin{align}
\bPsi^* &= \left( (\gamma+1) \I_M - \gamma^2 \A_{M,N} (\gamma \A_{N,N}^2 + \kappa \bar{\L} )^{-1} \A_{M,N}^\top
\right)^{-1} .
\end{align}
\end{small}\noindent
In words, $\bPsi^*$ is a function of interpolator $\bTheta$ and second denoiser $\bar{\bPsi}$; \ie, a new denoiser $\bPsi^*$ for $M$ original pixels is created due to the joint optimization \eqref{eq:MAP_mixed2}. 
The solution remains non-separable.

\subsection{Computation of Joint Denoising / Interpolation} 

Instead of computing joint operators $\bTheta^*$ and $\bPsi^*$ explicitly, one can compute the jointly denoised / interpolated output $\x^*$ by solving linear system $\C \x^* = \H^\top \y$ using \textit{conjugate gradient} (CG) \cite{Hestenes1952}. 
The complexity of CG solving a linear system $\A \x = \b$ is $\cO(\mathrm{nnz}(\A) \sqrt{\kappa(\A)} \log \frac{1}{\epsilon})$, where $\mathrm{nnz}(\A)$ is the number of non-zero entries in $\A$, $\kappa(\A)$ is the condition number, and $\epsilon$ is the convergence threshold.  
Substituting $\L$, $\bar{\L}$ and $\bTheta$ with their the corresponding denoiser and interpolator inverses via Theorem\;\ref{thm:denoiser} and \ref{thm:inter}, we get

\vspace{-0.1in}
\begin{footnotesize}
\begin{align}
\C = \left[ \begin{array}{cc}
(\gamma+1) \I_M + (\bPsi^{-1} - \I_M) & - \gamma \bTheta^{-1} \\
- \gamma \bTheta^{-\top} & \gamma \bTheta^{-\top} \bTheta^{-1} + (\bar{\bPsi}^{-1} - \I_N)
\end{array} \right] .
\label{eq:C_implement}
\end{align}
\end{footnotesize}\noindent
Note that $\mu$ and $\kappa$ are canceled out in \eqref{eq:C_implement}.
\eqref{eq:C_implement} is useful because it expresses the coefficient matrix directly in terms of available denoisers and interpolators, eliminating the need to explicitly construct graph Laplacian or adjacency matrices.
Solving linear system $\C \x^* = \H^\top \y$ via CG is as efficient as matrices $\bPsi^{-1}$, $\bTheta^{-1}$ and $\bar{\bPsi}^{-1}$ are sparse.

To mitigate computing matrix inverses $\bPsi^{-1}$, $\bTheta^{-1}$ and $\bar{\bPsi}^{-1}$ directly in \eqref{eq:C_implement}, with worst-case complexity $\cO(N^3)$, we adapt a sparse inverse estimation formulation called CLIME \cite{Cai2011_CLIME}, simultaneously benefiting from fast computation and resulting matrix sparsity that can speed up solution to the linear system $\C \x^* = \H^\top \y$.
CLIME is attractive because it directly estimates a sparse inverse through a convex optimization and admits column-wise parallelization.

\subsubsection{Computing $\L, \hat{\L}, \A$ given $\bPsi, \bar{\bPsi}, \bTheta$}

We approximate an inverse $\P$ of a matrix $\Q$ efficiently by adopting \textit{Constrained $\ell_1$-Minimization for Inverse Matrix Estimation} (CLIME) \cite{Cai2011_CLIME}.
CLIME minimizes the $\ell_1$-norm of matrix variable $\P$ so that $\Q \P$ is sufficiently close to an identity matrix $\I$, \ie,  
\begin{align}
\min_{\P} \|\P\|_1, 
~~ \mbox{s.t.}~ 
\|\Q \P - \I\|_{\infty} \leq \rho
\label{eq:CLIME}
\end{align}
where $\rho > 0$ is a constraint parameter. 
Note that \eqref{eq:CLIME} is always feasible for any $\rho \geq 0$ assuming $\Q$ is invertible.
Computing \eqref{eq:CLIME} for a chosen small value $\rho$ means that the approximate inverse $\P$ is in addition sparse.

Note that the columns $\{\p_i\}_{i=1}^N$ of $\P$ can be computed independently, resulting in $N$ optimizations running \textit{in parallel}:
\begin{align}
\min_{\p_i} \|\p_i\|_1, 
~~ \mbox{s.t.}~ 
\|\Q \p_i - \e_i\|_{\infty} \leq \rho
\label{eq:CLIME1}
\end{align}
where $\e_i$ is the $i$-th canonical vector---the $i$-th column of the identity matrix. 
\eqref{eq:CLIME1} is a \textit{linear program} (LP) and can be solved using a state-of-the-art general LP solver with complexity $\cO(N^{2.055})$ \cite{Jiang2021}. 

\subsubsection{Symmetrizing computed $\P$}

In general, the computed matrix $\P$---stacking obtained LP solutions $\p_i$'s as columns---is not symmetric.
When computing $\A_{M,N}$ given interpolator $\bTheta$ via Theorem\;\ref{thm:inter}, this is not a problem, since interpolator $\bTheta$ is not symmetric in general, and there is no expectation for $\A_{M,N}$ to be symmetric also.
On the other hand, when computing Laplacian $\L$ from symmetric denoiser $\bPsi$ (or $\bar{\L}$ from symmetric denoiser $\bar{\bPsi}$) via Theorem\;\ref{thm:denoiser}, we can first symmetrize computed $\P$ to $\tilde{\P}$, as done in \cite{Cai2011_CLIME}:  
\begin{align}
\tilde{P}_{i,j} = \begin{cases} P_{i,j}, & \mbox{if}~ |P_{i,j}| \le |P_{j,i}| \\ P_{j,i}, & \text{otherwise} \end{cases} .
\end{align}
\subsubsection{Diagonal Loading} Moreover, the symmetrized matrix $\tilde{\P}$ may not be PSD.
To ensure PSDness, we perform \textit{diagonal loading} using \textit{Gershgorin Circle Theorem} (GCT) \cite{varga04} as follows.
Define the center $c_i$ and radius $r_i$ of the $i$-th \textit{Gershgorin disc} of $\tilde{\P}$ as $c_i \triangleq \tilde{P}_{i,i}$ and $r_i \triangleq \sum_{j \neq i} |\tilde{P}_{i,j}|$, respectively. 
By GCT, the smallest disc left-end is a lower bound of the smallest eigenvalue $\lambda_{\min}$ of $\tilde{\P}$:
\begin{align}
\lambda^-_{\min}(\tilde{\P}) \triangleq \min_i c_i - r_i 
= \min_i \tilde{P}_{i,i} - \sum_{j \neq i} |\tilde{P}_{i,j}| .
\end{align}
Thus, to ensure PSDness, we augment $\tilde{\P}$ by $\tau \I$, where 
\begin{align}
\tau = \max \left( 0, 1 - \lambda^-_{\min}(\tilde{\P}) \right) .
\end{align}

%% file: interpolate.tex
One difficulty in using Theorem\;\ref{thm:inter} in practice is that interpolator $\bTheta$ must be square and full rank (and thus invertible).
This means that the number of new interpolated pixels $H$ is the same as the number of original pixels $M$, and the interpolated pixels must be linearly independent. 
In the case when the desired new pixels are fewer than the original pixels, \ie, $H < M$, this is not a problem:
we can artificially interpolate $M-H$ additional pixels that are independent, which are subsequently discarded after the interpolation operation\footnote{The $M-H$ additional interpolated pixels serve only to complete a full-rank square interpolator $\bTheta$ satisfying the assumptions of Theorem\;\ref{thm:inter}, and are discarded after interpolation.}.

We focus our discussion instead on the case when $H > M$. 
The key idea here is to partition the set of interpolated pixels into groups of $M$ linearly independent pixels. Each group is associated with an invertible interpolator $\bTheta_p$, allowing Theorem\;\ref{thm:inter} to be applied separately to each group.

Consider $P$ interpolators $\{\bTheta_p\}_{p=1}^P$, where each $\bTheta_p$ is of size $\mathbb{R}^{M \times M}$ and interpolates $M$ new pixels from $M$ original ones, for a total of $H =PM$ interpolated pixels. 
We can write the general linear interpolation operation for $\x \in \mathbb{R}^{(1+P)M}$ as
\begin{align}
\x = \left[ \begin{array}{c}
\I_M \\
\bTheta_1 \\
\vdots \\
\bTheta_P
\end{array} \right] \y .
\label{eq:interpolator2}
\end{align}

Next, denote by $\{\A_p\}_{p=1}^P$ the set of adjacency matrices corresponding to interpolators $\{\bTheta_p\}_{p=1}^P$, where each $\A_p \in \mathbb{R}^{(M+H) \times (M+H)}$ is defined as
\begin{align}
\A_p &= \left[ \begin{array}{cccc}
\0_{M,M} & \0_{M,H_p^-} & \A_{p,M,M} & \0_{M,H_p^+} \\
\0_{H,M} & \0_{H,H_p^-} & \0_{H,M} & \0_{H,H_p^+}
\end{array} \right] .  
\end{align}
$H_p^- \triangleq (p-1)M$ and $H_p^+ \triangleq (P-p)M$ are the number of interpolated pixels before and after the $p$-th interpolator, respectively.
Each (generally asymmetric) sub-matrix $\A_{p,M,M} \in \mathbb{R}^{M \times M}$ in $\A_p$ connects directional edges from $M$ new interpolated pixels to $M$ original pixels.

We generalize MAP objective \eqref{eq:MAP_inter} using GSV as prior to
\begin{align}
\min_{\x} \|\y - \H \x\|^2_2 + \gamma \sum_{p=1}^P \| \H (\x - \A_p \x)\|^2_2 .
\label{eq:MAP_inter2}
\end{align}
We now state the extension of Theorem\;\ref{thm:inter} for a single interpolator to multiple linear interpolators as follows.
\begin{theorem}
Interpolator $[\I_M; \bTheta_1; \ldots; \bTheta_P]$ is the solution filter to the MAP problem \eqref{eq:MAP_inter2} if $\A_{p,M,M} = \bTheta_p^{-1}, \forall p$, given all $P$ interpolators $\{\bTheta_p\}_{p=1}^P$ are square and full rank.
\label{thm:interpolator2}
\end{theorem}

\begin{proof}
Taking the derivative of the objective in \eqref{eq:MAP_inter2} w.r.t. $\x$ and setting it to $0$, we get

\vspace{-0.1in}
\begin{small}
\begin{align}
\left[ \H^\top \H + \gamma \sum_{p=1}^P  (\I_M - \A_p)^\top \H^\top \H (\I_M - \A_p) \right] \x^* = \H^\top \y  .
\end{align}
\end{small}\noindent
Using the definitions of $\H$ and $\A_p$, the linear system simplifies to $\C \x^* = \H^\top \y$, where coefficient matrix $\C$ is

\vspace{-0.05in}
\begin{footnotesize}
\begin{align}
\C = \left[ \begin{array}{cccc}
(1 + \gamma P) \I_M & -\gamma \A_{1,M,M} & -\gamma \A_{2,M,M} \ldots & -\gamma \A_{P,M,M} \\
-\gamma \A_{1,M,M}^\top & \gamma \A_{1,M,M}^2 & \0_{M,M} \ldots & \0_{M,M} \\
\vdots & & \ddots & \vdots \\
-\gamma \A_{P,M,M}^\top & \0_{M,M} & \ldots \0_{M,M} & \gamma \A_{P,M,M}^2
\end{array} \right] 
\end{align}
\end{footnotesize}\noindent 
where $\A^2_{p,M,M} = \A_{p,M,M}^\top \A_{p,M,M}$. 
Using matrix inverse formula \eqref{eq:inverse} and considering the lower-right sub-matrix $\mathrm{diag}(\gamma \A^2_{1,M,M}, \ldots, \gamma \A^2_{P,M,M})$ as the sub-block $\dot{\D}$, we first compute block $\dot{\P} = (\dot{\A} - \dot{\B} \dot{\D}^{-1} \dot{\C})^{-1}$:

\vspace{-0.1in}
\begin{small}
\begin{align}
\dot{\P} 
&= \left(
(1+\gamma P) \I_M - \gamma \sum_{p=1}^P \A_{p,M,M} (\A_{p,M,M}^2)^{-1} \A_{p,M,M}^\top
\right)^{-1}
\nonumber \\
&\stackrel{(a)}{=} \left( (1+\gamma P) \I_M - \gamma P \I_M \right)^{-1} ~=~ \I_M .
\end{align}
\end{small}\noindent
In $(a)$, we apply the assumption that $\A_{p,M,M} = (\bTheta_p)^{-1}, \forall p$.

Using \eqref{eq:inverse} and $\dot{\P} = \I_M$, we can now write $\x^*$ as

\vspace{-0.1in}
\begin{small}
\begin{align}
\x^* &= \C^{-1} \H^\top \y 
\nonumber \\
&= \left[ \begin{array}{c}
\I_M \\
- \mathrm{diag}^{-1}(\gamma \A_{p,M,M}^2\}) 
~\text{vec}(\{-\gamma \A_{p,M,M}^\top\}) 
\end{array} \right] \y
\nonumber \\
&= \left[ \begin{array}{c}
 \I_M \\
(\A^2_{1,M,M})^{-1} \A_{1,M,M}^\top \\
\vdots \\
(\A^2_{P,M,M})^{-1} \A_{P,M,M}^\top 
\end{array} \right] \y 
~\stackrel{(a)}{=}~ \left[ \begin{array}{c}
\I_M \\
\bTheta_1 \\
\vdots \\
\bTheta_P
\end{array} \right] \y ,
\end{align}
\end{small}\noindent
where in $(a)$ we apply the assumption that $\A_{p,M,M} = \bTheta_p^{-1}, \forall p$. 
\end{proof}

\noindent
\textbf{Remark}: Theorem \ref{thm:interpolator2} shows that a collection of invertible interpolators can be represented by a collection of directed graph smoothness priors, extending Theorem \ref{thm:inter} from a single interpolation block to multiple interpolation blocks.

\subsection{Computation of Joint Denoising / Multi-Interpolation}

The joint denoising / interpolation formulation \eqref{eq:MAP_mixed2} generalizes for the multiple-interpolator case to

\vspace{-0.05in}
\begin{small}
\begin{align}
\min_{\x} & \| \y - \H \x \|^2_2 + \gamma \sum_{p=1}^P \| \H (\x - \A_p \x) \|^2_2 + \mu (\H \x)^\top \L (\H \x) 
\nonumber \\
& + \kappa (\G \x)^\top \bar{\L} (\G \x) .
\label{eq:obj_mixed_mod3}
\end{align}
\end{small}\noindent
We show how \eqref{eq:obj_mixed_mod3} can be solved efficiently.
Generalizing from \eqref{eq:C_implement}, solution $\x^*$ to \eqref{eq:obj_mixed_mod3} can be obtained by solving a linear system $\C \x^* = \H^\top \y$, where the coefficient matrix $\C$ is

\vspace{-0.1in}
\begin{footnotesize}
\begin{align}
\C &= \left[ \begin{array}{c|c}
(1 + \gamma P) \I_M + (\bPsi^{-1} - \I_M) & -\gamma \bTheta_1^{-1} \ldots -\gamma \bTheta_P^{-1}   \\ \hline
\begin{array}{c}
-\gamma \bTheta_1^{-\top} \\
\vdots \\
-\gamma \bTheta_P^{-\top}
\end{array}
     & \D + (\bar{\bPsi}^{-1} - \I_{PM})
\end{array} \right] 
\nonumber \\
\D &= \gamma \mathrm{diag}(\bTheta_1^{-\top} \bTheta_1^{-1}, \ldots, \bTheta_P^{-\top} \bTheta_P^{-1}) .
\end{align}
\end{footnotesize}\noindent

As done previously, linear system $\C \x^* = \H^\top \y$ can be solved efficiently via CG.

\subsection{Application of Theorem\;\ref{thm:interpolator2} for Rectangular $\bTheta$}
\label{subsec:app_rectangular}

When given a general interpolator matrix $\bTheta$ that is rectangular, we use Theorem\;\ref{thm:interpolator2} as follows.
If $\bTheta \in \mathbb{R}^{H \times M}$ is a \textit{long} matrix (assuming the rows are independent) for $H < M$, then we simply add $M-H$ linearly independent \textit{dummy} rows to $\bTheta$, so that the resulting interpolator is square.
Any rows that preserve full rank may be used.

On the other hand, if $\bTheta \in \mathbb{R}^{H \times M}$ is a \textit{tall} matrix (assuming the columns are independent) for $H > M$, then we first divide $\bTheta$ into $P = \lfloor \frac{H}{M} \rfloor$ square interpolators $\bTheta_p$, each of size $M \times M$.
Then, to the remaining $H - PM < M$ rows in $\bTheta$, we add $M - (H - PM)$ dummy rows to compose $\bTheta_{P+1}$; \ie, 
\begin{align}
\bTheta' = \left[ 
\begin{array}{c}
\bTheta_1 \\
\vdots \\
\bTheta_P \\
\bTheta_{P+1} 
\end{array}
\right] .
\end{align}
We can now treat modified $\bTheta'$ as a set of $P+1$ interpolators $\{\bTheta_i\}_{i=1}^{P+1}$, each of size $M \times M$.

%% file: results.tex
\subsection{Experimental Setup}

We conducted extensive experiments to test our derived operators in \eqref{eq:MAP_mixed2_linSys2} for joint image denoising / interpolation (denoted by \texttt{joint}) and compared them against original sequential operations (denoted by \texttt{sequential}) in the non-separable case. 
As stressed in the Introduction, our goal is not to design new denoisers or interpolators, but to optimally combine existing ones within a graph-theoretic framework (Fig.~\ref{fig:joint_overview}). Accordingly, competing schemes are restricted to different combinations of the same denoisers and interpolators.

\vspace{0.05in}
\noindent
\textbf{Baseline Denoisers:}
For denoisers, we chose well-known and representative denoisers in the literature: Gaussian filter \cite{milanfar13}, \textit{bilateral filter} (BF) \cite{tomasi98}, and \textit{nonlocal means} (NLM) \cite{buades05}. 
The Gaussian filter is \textit{linear}: its positive weights depend only on the differences between pixel locations.
In contrast, BF and NLM are \textit{pseudo-linear}: the denoiser $\bPsi(\x)$ depends on the current estimate $\x_t$, and is treated as a fixed linear operator within each iteration.
Specifically, each BF weight is a product of domain and range filters, where the domain filter weights depend on the differences between pixel locations, while the range filter weights depend on the differences between pixel intensities.
On the other hand, NLM first identifies similar patches in an image via a restricted search, and computes filter weights based on the differences between pixel intensities in a mean patch computed from identified similar patches.

To ensure an input denoiser $\bPsi$ to our optimization framework was non-expansive, symmetric and PD according to Theorem\;\ref{thm:denoiser}, given a symmetric, non-negative affinity matrix $\bPsi^o$ constructed from Gaussian, BF, or NLM weights (prior to row normalization), we did the following.
First, we performed \textit{symmetric normalization} $\tilde{\bPsi} = \D^{-1/2} \bPsi^o \D^{-1/2}$, where $\D$ is the diagonal degree matrix with $D_{i,i} = \sum_{j} \Psi^o_{i,j}$, so that $\tilde{\bPsi}$ was symmetric with spectral norm bounded by $1$.
Second, we perturbed $\tilde{\bPsi}$ to $\bPsi = (1-\epsilon) \tilde{\bPsi} + \epsilon \I$ for small $\epsilon > 0$ ($\epsilon = 10^{-3}$ in all experiments), so that $\bPsi$ was strictly PD with eigenvalues in range $[\epsilon, 1]$.
This convex combination preserved non-expansiveness.
This construction ensured that $\bPsi$ satisfied the conditions in Theorem~\ref{thm:denoiser}.

\vspace{0.05in}
\noindent
\textbf{Baseline Interpolators:}
For interpolators, we used linear operators for image rotation, warping via homography transform, and demosaicking via bilinear interpolation. 

To ensure that the numbers of input pixels and output pixels are equal (\ie, $M = N$), we augmented $\bTheta$ with additional rows corresponding to dummy pixels, chosen to be linearly independent.
It is important to ensure that $\bTheta$ remains full rank (thus invertible) when adding new rows. 
Further, to improve numerical stability, we added a scaled identity matrix, $\epsilon \I$ for small $\epsilon > 0$, to $\bTheta$ if the \textit{condition number} $\kappa(\bTheta)$ of matrix $\bTheta$ was too large.
$\kappa(\bTheta)$ is defined as
\begin{align}
\kappa(\bTheta) \triangleq \frac{\sigma_{\max}(\bTheta)}{\sigma_{\min}(\bTheta)} ,
\end{align}
which is the ratio of the largest to smallest singular values of $\bTheta$. 
$\kappa(\bTheta)$ can be estimated efficiently using 1-norm instead, without full singular value decomposition (SVD) or computing $\bTheta^{-1}$, as $\tilde{\kappa}(\bTheta) = \|\bTheta\|_1 \widetilde{\|\bTheta^{-1}\|_1}$, where $\widetilde{\|\bTheta^{-1}\|_1}$ can be estimated using the Hager-Higham algorithm \cite{Hager1984,Higham2000}. 
Summarizing, if $\tilde{\kappa}(\bTheta) > 10^3$, then we add $\epsilon \I$, where $\epsilon$ is
\begin{align}
\epsilon = \frac{\|\bTheta\|_1}{10^3} .
\end{align}

\vspace{0.05in}
\noindent
\textbf{Experimental Details:}
iid Gaussian noise of different variances were added to test images.
The output images were evaluated using peak signal-to-noise ratio (PSNR), as conventionally done in the image compression / restoration literature. 
Demosaicking was evaluated on a $496 \times 496$ Stained-glass image. 
Rotation and warping were evaluated on the $512 \times 512$ grayscale Peppers image. 
For rotation with the Gaussian and BF denoisers, we also repeated the experiments on the $512 \times 512$ grayscale Airplane image, to check that the observed trends do not depend on the test image.

The joint denoising / interpolation operation was performed targeting $8 \times 8$ output patches for demosaicking and rotation, and $5 \times 5$ for warping.
The size and location of the input patch from which the output patch was generated depend on the interpolation operation and the output location.

Note that for a chosen interpolation operation, $\bTheta$ changed from patch to patch, and so the denoiser must adapt to the input and output dimensions when using the Gaussian filter. 
For BF, and NLM, the denoiser itself changed from patch to patch. 
Each figure shows \texttt{joint} computed by solving the linear system $\C \x^* = \H^\top \y$ via CG, as well as exact matrix inverses (no CLIME), so that the accuracy cost of the sparse CLIME estimates can be evaluated. 
Results are shown for 1 and 2 denoiser passes.

\begin{figure}[h]
    \centering
    \includegraphics[width=\linewidth]{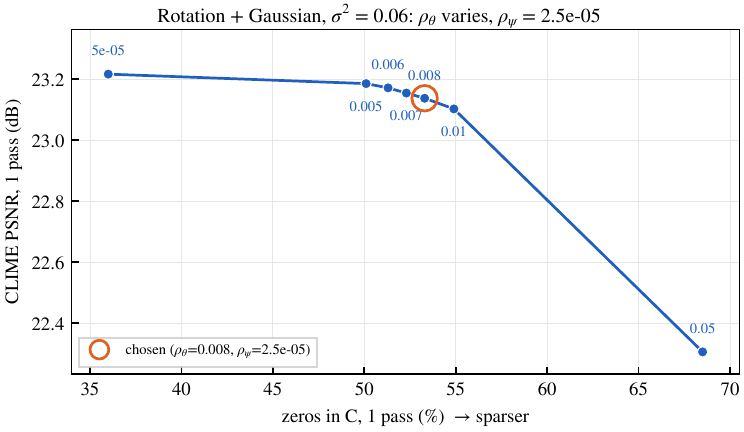}
    \vspace{-0.2in}
    \caption{Effect of $\rho_\theta$ on the sparsity of $\mathbf{C}$ and on PSNR (rotation + Gaussian, $\sigma^2 = 0.06$, 1 pass, $\rho_\psi = 2.5\times10^{-5}$). Labels give $\rho_\theta$; the circle marks the chosen value.}
    \label{fig:rho_theta_study}
\end{figure}

The CLIME thresholds $\rho_\theta$ and $\rho_\psi$ control how sparse the estimated matrix inverses are for interpoloators an denoisers, respectively.
Larger values yield a sparser $\mathbf{C}$, and thus cheaper CG iterations, at the cost of accuracy (Fig.~\ref{fig:rho_theta_study}).
For each experiment, we chose $(\rho_\theta, \rho_\psi)$ at the knee of this trade-off; for example, for rotation + Gaussian, the number of zeros in $\mathbf{C}$ increase from $35.5\%$ (exact inverse) to $53.3\%$ with only a $0.08$\,dB PSNR loss.

\subsection{Results for Joint Denoising / Demosaicking}

We first applied our proposed optimization framework to the joint denoising / demosaicking problem, where the goal was to reconstruct a full-resolution image from a noisy Bayer-pattern observation. 
A $496 \times 496$  Stained-glass image was used for testing, with iid Gaussian noise of variance $\sigma^2 \in \{0.02, 0.04, 0.06, 0.08, 0.1\}$ added to the raw Bayer observations prior to demosaicking.

\vspace{0.05in}
\noindent
\textbf{Demosaicking Interpolator:} 
Interpolator $\bTheta$ was constructed to implement bilinear Bayer-pattern interpolation for the blue channel. 
For each $8 \times 8$ output patch ($N = 64$ pixels), the observed blue pixels occupied every other row and column, yielding $M = 16$ input observations. Non-blue pixel positions were assigned bilinear averaging weights from their nearest blue neighbors. The resulting matrix was of size $64 \times 16$. 
Tall matrix $\bTheta$ was vertically partitioned into $P = N/M = 4$ sub-interpolators, and matrix $\C$ was assembled by iterating over the four $16 \times 16$ sub-blocks of $\bTheta$.

\vspace{0.05in}
\noindent
\textbf{Denoiser Construction:} A Gaussian filter with standard deviation $\sigma = 0.4$ and kernel half-width of 1 was used to construct both the source denoiser $\bPsi$ (on the $4 \times 4$ blue-pixel domain) and target denoiser $\bar{\bPsi}$ (on the full $8 \times 8$ output patch). 
Since patch dimensions were fixed across all patches, both denoiser matrices and their matrix powers are precomputed once before the processing loop, reducing per-patch overhead.

\vspace{0.05in}
\noindent
\textbf{Implementation Details:} 
The CLIME regularization parameters were set to $\rho_\theta = 3 \times 10^{-2}$ for the interpolator and $\rho_\psi = 5 \times 10^{-5}$ for the denoiser. 
The GSV weight was set to $\gamma = 0.2$. 
The linear system $\C \x^* = \H^\top \y$ was solved using CG with a tolerance of $10^{-3}$.

\begin{figure}[t]
    \centering
    \includegraphics[width=\linewidth]{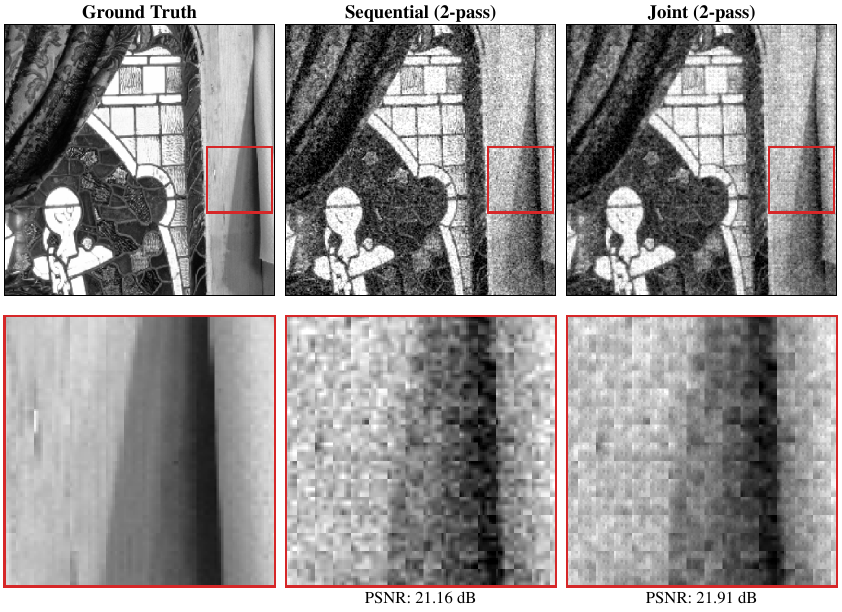}
    \caption{Visual comparison of Sequential vs.\ Joint 2-pass demosaicking with Gaussian
             denoising (noise variance = 0.02). Top: full demosaicked output with highlighted
             region of interest. Bottom: zoomed crop showing reconstruction detail.}
    \label{fig:demosaic_gaus_comparison}
\end{figure}

\begin{figure}[h]
    \centering
    \includegraphics[width=\linewidth]{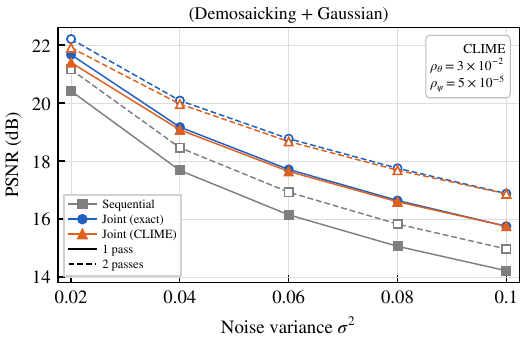}
    \vspace{-0.25in}
    \caption{Demosaicking + Gaussian with CLIME}
    \label{fig:demosaic_gaus_plot}
\end{figure}

\vspace{0.05in}
\noindent
\textbf{Baseline Comparison:} 
The sequential baseline applied the source denoiser to the observed blue pixels, interpolated to the full grid, and applied the target denoiser: $\bar{\bPsi} \bTheta \bPsi \mathbf{x}_\text{blue}$.
As discussed in \cite{milanfar13}, image filters are often applied in multiple passes for more effective performance.
Thus, multi-pass results ($n = 1, 2$) were evaluated for both methods by applying matrix powers of the precomputed denoiser matrices. 
The ground truth for PSNR evaluation was obtained by applying $\bTheta$ to the noise-free blue channel. Fig.\;\ref{fig:demosaic_gaus_comparison} shows a visual comparison at $\sigma^2 = 0.02$ where we observe that the \texttt{joint} recovers the image better than \texttt{sequential}, especially in the curtain region. 
Fig.\;\ref{fig:demosaic_gaus_plot} shows that \texttt{joint} performs better than \texttt{sequential} at all noise levels for both 1 and 2 denoiser passes, while staying within $0.3$\,dB of \texttt{joint} with exact inverses.
The gap increases with noise level in all scenarios, with maximum PSNR gain of $1.9$dB observed at noise variance $0.1$ in the 2-denoiser-pass experiment.

\subsection{Results for Joint Denoising / Rotation}
\label{subsec:denoiser_rot_exp}

To maintain computational tractability, the input images were processed in non-overlapping $8 \times 8$ patches. 
To ensure the invertibility of rectangular interpolator matrices encountered during patch rotation (where the number of source and target pixels differ), we applied the augmentation strategy described in Section\;\ref{subsec:app_rectangular}. 
The resulting linear system \eqref{eq:MAP_mixed2_linSys} was solved using CG with a convergence tolerance of $10^{-6}$.

\begin{figure}[h]
    \centering
    \includegraphics[width=1.02\linewidth]{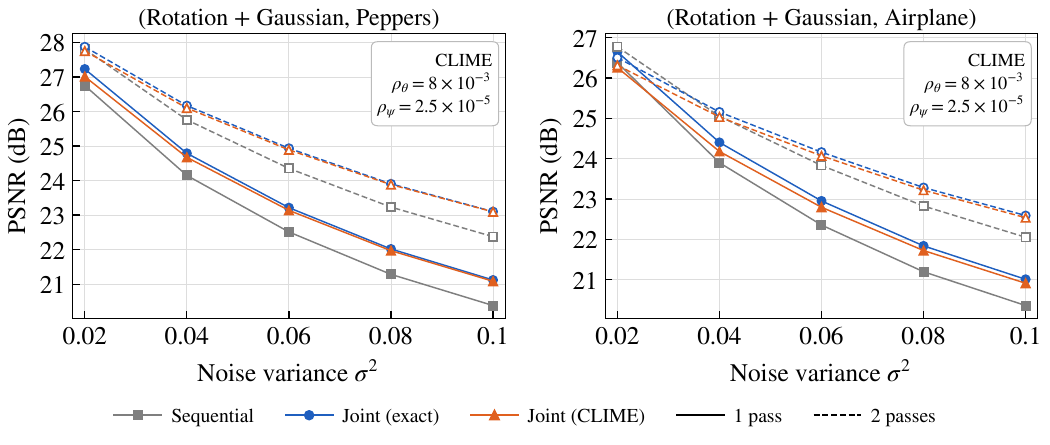}
    \vspace{-0.2in}
    \caption{Rotation + Gaussian with CLIME for Peppers and Airplane images, respectively.}
    \label{fig:rot_gaus_plot}
\end{figure}

The GSV weight was set to $\gamma = 0.5$.
The Gaussian denoiser utilized a standard deviation of $\sigma=0.6$ with a $3 \times 3$ kernel, and the CLIME thresholds were $\rho_\theta = 8 \times 10^{-3}$ and $\rho_\psi = 2.5 \times 10^{-5}$.
Fig.\;\ref{fig:rot_gaus_plot} shows the performance of \texttt{joint} vs \texttt{sequential} for two test images under multiple denoiser passes. 
In general, \texttt{joint} (orange) performs better than \texttt{sequential} (grey) and stays within $0.22$\,dB of \texttt{joint} with exact inverses (blue).
Maximum PSNR gain of $0.72$\,dB was observed for the 2-denoiser-pass at noise variance $0.1$.

In the second set of experiments, we evaluated the framework's flexibility by employing a pesudo-linear BF as the base denoiser. 
Unlike the Gaussian filter, BF is known to preserve image structure such as sharp object boundaries.

\vspace{0.05in}
\noindent
\textbf{Graph Construction for Signal-Dependent Filters:} 
To employ the signal-dependent BF in our framework, we constructed the weight matrices dynamically for each $8 \times 8$ patch. 
The source denoiser matrix $\bPsi$ was first computed using the noisy source patch. 
For the target denoiser $\bar{\bPsi}$, we need a rough estimate of the output signal which was calculated as $\bTheta \y$.

The BF weights were calculated using a spatial standard deviation $\sigma_s = 0.7$ and a range standard deviation $\sigma_r = 0.7$ with a kernel width of 2. 
These weight matrices were then normalized and perturbed as described in the experimental setup, so that their eigenvalues lie in $[\epsilon, 1]$.

\vspace{0.05in}
\noindent
\textbf{Baseline Comparison:} The sequential baseline for this experiment was defined as a three-stage cascade: $\bar{\bPsi} \bTheta \bPsi \y$. 
The noisy input is first filtered by the source BF denoiser $\bPsi$, rotated via $\bTheta$, and finally filtered by the target BF denoiser $\bar{\bPsi}$.
The GSV weight was set to $\gamma = 0.5$, with noise variances tested across the range $\sigma^2 \in [0.02, 0.1]$, and CLIME thresholds $\rho_\theta = 10^{-2}$ and $\rho_\psi = 5 \times 10^{-5}$.

\begin{figure}[h]
    \centering
    \includegraphics[width=1.02\linewidth]{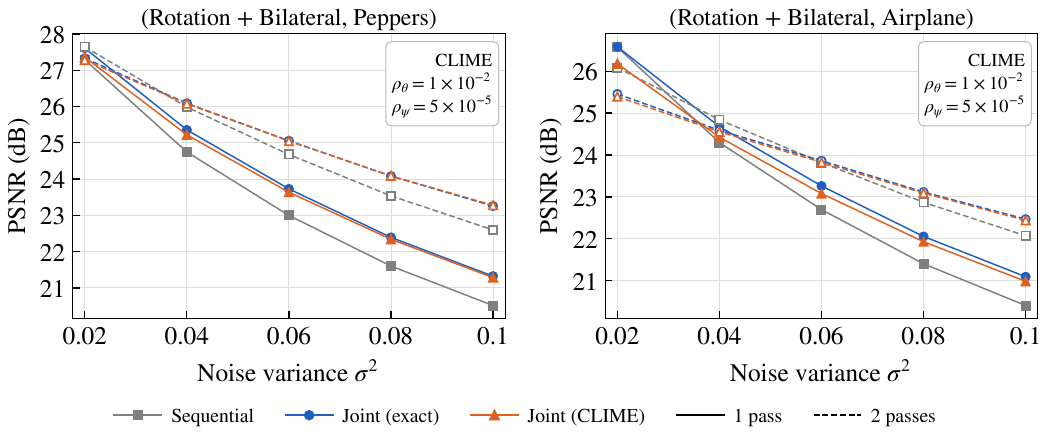}
    \vspace{-0.25in}
    \caption{Rotation + Bilateral with CLIME for Peppers and Airplane images, respectively.}
    \label{fig:rot_bil_plot}
\end{figure}

Fig.\;\ref{fig:rot_bil_plot} shows a similar trend to Fig. \;\ref{fig:rot_gaus_plot} where \texttt{joint} outperforms \texttt{sequential} generally at higher noise levels. 
Maximum PSNR gain of $0.77$dB is observed at noise variance $0.1$ in the 1-denoiser-pass.
Due to the similar performance trends observed in the two test images, we focus on the Pepper test image in the sequel.

\begin{figure}[h]
    \centering
    \includegraphics[width=\linewidth]{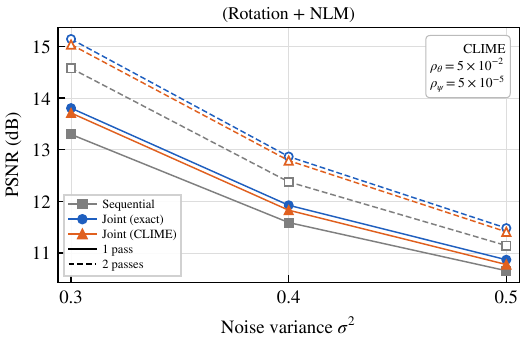}
    \vspace{-0.25in}
    \caption{Rotation + NLM with CLIME}
    \label{fig:rot_nlm_plot}
\end{figure}

Finally, we used NLM as the denoiser. 
Its weights were computed as $\exp(-\|\mathbf{p}_i - \mathbf{p}_j\|_2^2 / h)$ with $h = 0.4$, where $\mathbf{p}_i$ is the $3 \times 3$ neighborhood of pixel $i$, over all pixel pairs within the patch; $\bPsi$ was computed from the noisy source patch and $\bar{\bPsi}$ from $\bTheta \y$.
We used noise variances $\sigma^2 \in [0.3, 0.5]$ and CLIME thresholds $\rho_\theta = 5 \times 10^{-2}$ and $\rho_\psi = 5 \times 10^{-5}$; the remaining settings were the same as the Gaussian denoiser.
Fig.\;\ref{fig:rot_nlm_plot} shows that \texttt{joint} outperforms \texttt{sequential} at all tested noise levels, with a maximum gain of $0.46$\,dB at noise variance $0.3$ in the 2-denoiser-pass.

\subsection{Results for Joint Denoising / Warping }

\begin{figure}[t]
    \centering
    \includegraphics[width=\linewidth]{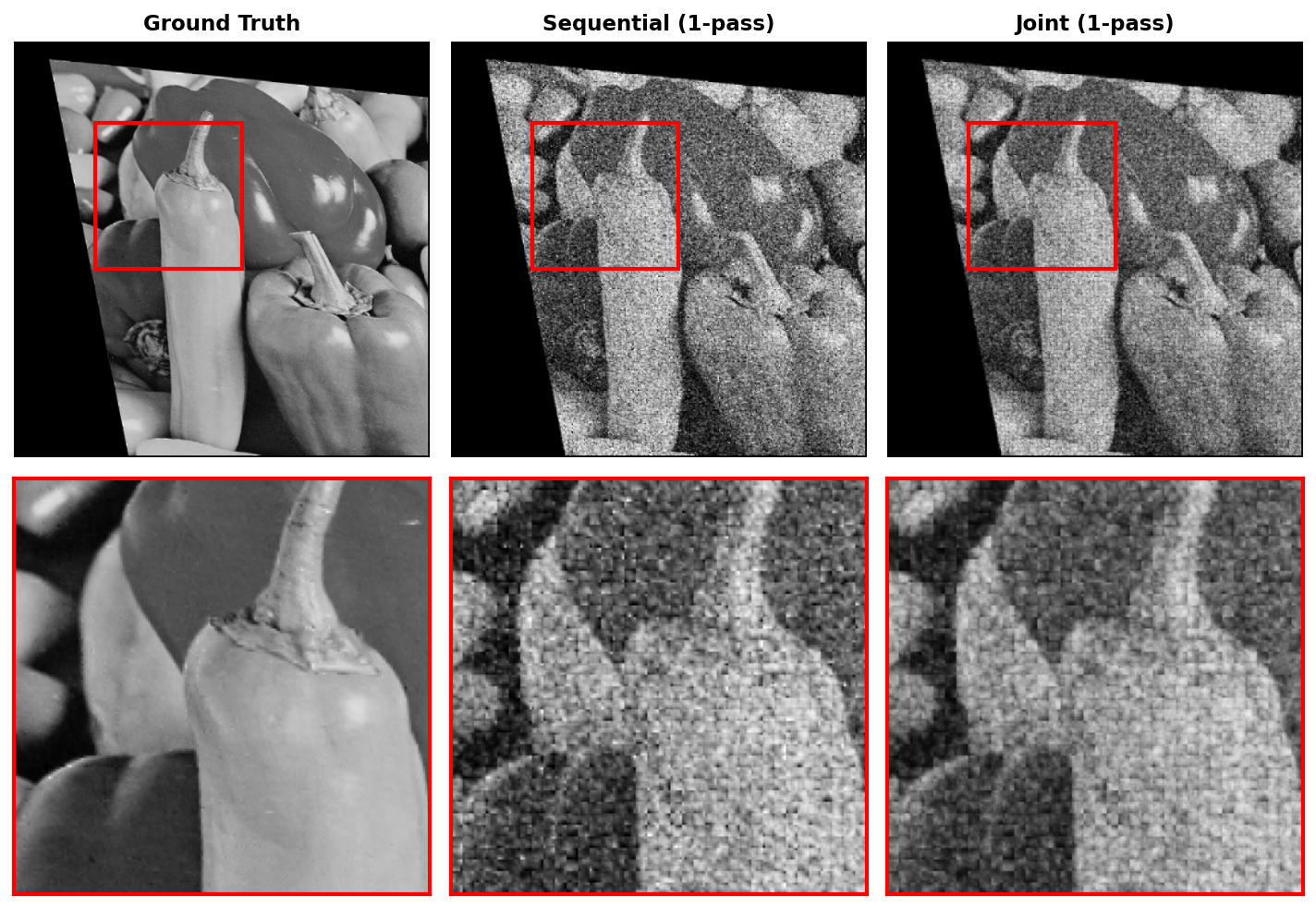}
    \caption{Visual comparison of Sequential vs.\ Joint 1-pass warping with bilateral denoising
             (noise variance = 0.08). Top: full warped output with highlighted region of interest.
             Bottom: zoomed crop showing noise suppression detail.}
    \label{fig:warp_bilateral_comparison}
\end{figure}

To evaluate the framework's robustness against complex geometric deformations, we conducted experiments involving image warping defined by a homography. 
Unlike simple rotation, homography introduces non-uniform pixel displacements and perspective distortions, making the interpolation irregular.

\vspace{0.05in}
\noindent
\textbf{Homography and Backward Mapping:} The geometric transformation was defined by the homography matrix $\Q = [[1, 0.2, 0], [0.1, 1, 0], [0, 0, 1]]$, which introduces both shear and perspective effects. 
We used a backward mapping strategy: for each $5 \times 5$ target patch, we computed the corresponding source coordinates $(u, v)$ using the inverse homography. 
These continuous coordinates were used to populate the interpolator matrix $\bTheta$ via bilinear interpolation weights.

We employed a Gaussian denoiser ($\sigma=0.6$, kernel size $5 \times 5$) for both the source and target signals. 
The joint optimization problem was solved on $5 \times 5$ patches to handle the spatial variability of the warp. The GSV weight was set to $\gamma=0.5$. The CLIME sparsity parameters were set to $\rho_{\theta}=2.4 \times 10^{-2}$ and $\rho_{\psi}=4 \times 10^{-3}$ .

\begin{figure}[h]
    \centering
    \includegraphics[width=\linewidth]{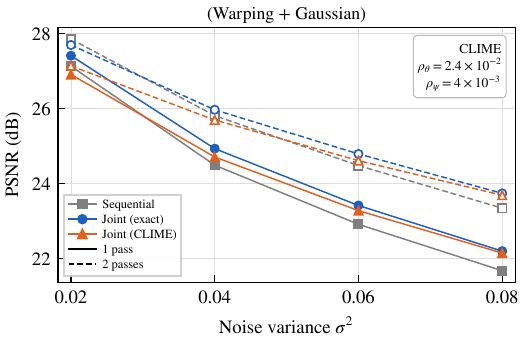}
    \vspace{-0.25in}
    \caption{Warping + Gaussian with CLIME}
    \label{fig:warp_gaus_plot}
\end{figure}

\vspace{0.05in}
\noindent
\textbf{Baseline Comparison}:
Fig.\;\ref{fig:warp_gaus_plot} shows that the \texttt{joint} performs better than \texttt{sequential} for $\sigma^2 \geq 0.04$ with 1 pass and $\sigma^2 \geq 0.06$ with 2 passes, with increasing gap as noise increases.
Maximum PSNR gain of $0.47$\,dB was observed at noise variance $0.08$ in the 1-denoiser-pass experiment.

In the next set of experiments, we considered joint restoration under geometric deformation (homography) using signal-dependent BF for denoising. 
This combination tests the framework's ability to simultaneously handle irregular graph topologies induced by the warp and content-adaptive edge weights induced by the denoiser.

\vspace{0.05in}
\noindent
\textbf{Signal-Dependent Graph Construction on Irregular Grids:} The homography was defined as $\Q = [[1, 0.2, 0], [0.1, 1, 0], [0, 0, 1]]$, introducing shear and perspective distortions. 
For the source domain, the BF weights ($\bPsi$) were computed directly from the noisy input patch. 
For the target domain, the weights ($\bar{\bPsi}$) were constructed using the warped signal structure.
The BF parameters were set to $\sigma_s = 0.6$ (spatial) and $\sigma_r = 0.6$ (range) with a kernel width of 2, and the CLIME thresholds were $\rho_\theta = 2.2 \times 10^{-2}$ and $\rho_\psi = 3 \times 10^{-3}$.
The optimization was performed on $5 \times 5$ patches with GSV weight $\gamma = 0.5$. 

\begin{figure}[h]
    \centering
    \includegraphics[width=\linewidth]{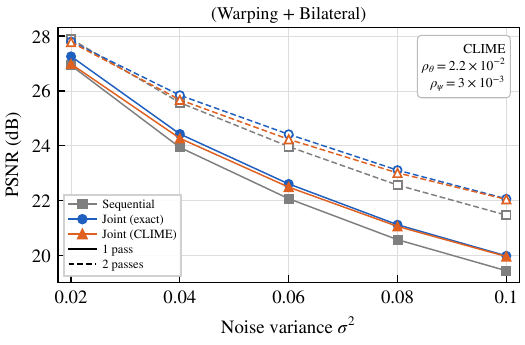}
    \vspace{-0.25in}
    \caption{Warping + Bilateral with CLIME}
    \label{fig:warp_bil_plot}
\end{figure}

\vspace{0.05in}
\noindent
\textbf{Baseline Comparison:} 
The sequential baseline was defined as the cascade $\bar{\bPsi} \bTheta \bPsi \y$.
Similar to Fig. \;\ref{fig:warp_gaus_plot}, we see that with BF in Fig. \;\ref{fig:warp_bil_plot} \texttt{joint} outperforms \texttt{sequential} at all noise levels, except at $\sigma^2 = 0.02$ with 2 passes.
Maximum PSNR gain of $0.57$\,dB was observed at noise variance 0.1 in the 2-denoiser-pass experiment.
We also observe how the performance is visually compelling in Fig.\;\ref{fig:warp_bilateral_comparison}.
We see that the skin region is better recovered using \texttt{joint}.

%% file: conclude.tex
We proposed a mixed-graph filtering framework to combine existing pseudo-linear denoisers and interpolators into jointly optimal image denoising/interpolation operators. 
We first established graph-theoretic MAP interpretations of denoisers and interpolators, showing that under suitable conditions a pseudo-linear denoiser corresponds to an undirected graph filter regularized by the graph Laplacian regularizer (GLR), while a pseudo-linear interpolator corresponds to a directed graph filter regularized by graph shift variation (GSV). 
These equivalence results enable denoising and interpolation operators to be incorporated into a common optimization framework. 

Building on this foundation, we developed mixed directed-undirected graph formulations for joint image denoising and interpolation. 
We showed that a basic formulation yields a separable optimum corresponding to the conventional denoise-then-interpolate pipeline, while introducing graph regularization on interpolated pixels leads to a non-separable optimum with jointly optimal denoising and interpolation operators. 
To enable practical deployment, we developed an efficient implementation based on sparse inverse estimation and iterative linear-system solvers. 

Experimental results demonstrated consistent improvements over separate denoising and interpolation pipelines across multiple denoisers, interpolators, and imaging tasks. The proposed framework provides a principled graph-theoretic approach for transforming existing denoisers and interpolators into jointly optimized reconstruction operators.

\section*{Acknowledgment}
The authors used OpenAI GPT-6 for language editing and grammar polishing of the manuscript.